\documentclass[superscriptaddress,twocolumn,
 amssymb,amsmath,nobibnotes,aps,prd,showkeys,showpacs,nofootinbib]{revtex4} 
\usepackage{graphicx}
\usepackage{enumerate}
\usepackage{hyperref}
\newcommand{\be}{\begin{equation}}
\newcommand{\ee}{\end{equation}}
\newcommand{\bear}{\begin{eqnarray}}
\newcommand{\eear}{\end{eqnarray}}
\newcommand{\m}{\phantom{$-$}}

\begin{document}

\begin{flushleft}
KCL-PH-TH/2011-23 \\
CERN-PH-TH/2011-160 \\
LCTS/2011-09 \\
IFIC/11-36
\end{flushleft}

\vspace{0.4cm}

\title{Dynamics and constraints of the dissipative Liouville cosmology}

\author{Spyros Basilakos}
\affiliation{Academy of Athens, Research Center for Astronomy and Applied Mathematics,
 Soranou Efesiou~4, 11527, Athens, Greece}
\affiliation{High Energy Physics Group, Dept.\ ECM, Universitat de Barcelona,
Av.\ Diagonal 647, E-08028 Barcelona, Spain}

\author{Nick E.\  Mavromatos}
\affiliation{King's College London, Department of Physics, Theoretical Physics, London WC2R~2LS, UK}
\affiliation{CERN, Theory Division, Physics Department, CH-1211 Geneva 23, Switzerland}

\author{Vasiliki A.\ Mitsou\footnote{Corresponding author. \\E-mail address: vasiliki.mitsou@ific.uv.es\\ Tel: +34963543855; fax: +34 963543488}}
\affiliation{Instituto de F\'isica Corpuscular (IFIC), CSIC -- Universitat
  de Val\`encia, P.O.~Box~22085, E-46071, Valencia, Spain}

\author{Manolis Plionis}
\affiliation{Institute of Astronomy \& Astrophysics, National Observatory of Athens,
Thessio 11810, Athens, Greece }
\affiliation{Instituto Nacional de Astrof\'isica, \'Optica y Electr\'onica, 72000 Puebla, Mexico}

\begin{abstract}
In this article we investigate the properties of the FLRW flat cosmological models
in which the cosmic expansion of the Universe is affected by
a dilaton dark energy (Liouville scenario). In particular,
we perform a detailed study of these models
in the light of the latest cosmological data, which serves to
illustrate the phenomenological viability of the new dark energy
paradigm as a serious alternative to the traditional scalar field
approaches. By performing a joint likelihood analysis of the recent
supernovae type~Ia data (SNIa),
the differential ages of passively evolving galaxies, and the
Baryonic Acoustic Oscillations (BAOs)
traced by the Sloan Digital Sky Survey (SDSS), we put tight
constraints on the main cosmological parameters.
Furthermore, we study the linear matter fluctuation field
of the above Liouville cosmological models.
In this framework, we compare the observed growth rate of
clustering measured 
with those predicted
by the current Liouville models. Performing a $\chi^{2}$ statistical test
we show that the Liouville cosmological model
provides growth rates that match sufficiently well with the
observed growth rate.
To further test the viability of the models under study, we use
  the Press-Schechter formalism to derive their expected
redshift distribution of cluster-size halos that will be provided by
future X-ray and Sunyaev-Zeldovich
cluster surveys. We find that the Hubble flow differences
between the Liouville and the $\Lambda$CDM models provide
a significantly different halo redshift distribution, suggesting
that the models can be observationally distinguished.

\end{abstract}
\maketitle

\section{Introduction}\label{sec:intro}
The comprehensive study
carried out in recent years by the cosmologists has
converged towards a cosmic expansion history that
involves a spatially flat geometry and a recent accelerating
expansion of the Universe (see
\cite{Teg04,Spergel07,essence,Kowal08,Hic09,komatsu08,LJC09,BasPli10}
and references therein).
From a theoretical point of view, an easy way to explain this
expansion is to consider an additional energy component, usually called dark
energy (hereafter DE) with negative pressure, that
dominates the Universe at late times.
The simplest DE candidate corresponds to the cosmological constant (see
\cite{Weinberg89,Peebles03,Pad03} for reviews). Indeed the so-called spatially
flat concordance $\Lambda$CDM model, which includes cold dark matter (CDM) and
a cosmological constant ($\Lambda$), fits accurately the current observational
data and thus it is an excellent candidate to be the model which describes the
observed Universe.

Nevertheless, the identification of $\Lambda$ with the quantum
vacuum has brought another problem: the estimate of theoretical
physicists that the vacuum energy density should be 120 orders of
magnitude bigger than the measured $\Lambda$ value. This is the
``old" cosmological constant problem (CCP) \cite{Weinberg89}. The
``new" problem \cite{Peebles03} asks why is the vacuum density so
similar to the matter density just now? Many solutions to both
problems have been proposed in the literature
\cite{Zlatev99,AASW09,Egan08}.
An easy way to overpass the above theoretical problems is to replace the
constant vacuum energy with a DE
that evolves with time. Nowadays, the physics
of DE is considered one of the most fundamental and challenging problems on
the interface uniting Astronomy, Cosmology and Particle Physics and indeed
in the last
decade there have been theoretical debates among the cosmologists regarding
the nature of this exotic component.

In the original scalar field models~\cite{Dolgov82} and later in the
quintessence context, one can ad-hoc introduce an adjusting or tracker scalar
field $\phi$~\cite{Caldwell98} (different from the usual SM Higgs field),
rolling down the potential energy $V(\phi)$, which could resemble the DE~\cite{Peebles03,Pad03,Jassal,SR,Xin,SVJ}. This class
of DE models have been widely used in the literature due to their simplicity.
Notice that DE models with a canonical kinetic term have a dark energy EoS
parameter $-1\leq w_{\phi}<-1/3$. Models with ($w_{\phi}<-1$), sometimes
called phantom DE~\cite{phantom}, are endowed with a very exotic nature, like
a scalar field with negative kinetic energy.
However, it was realized that the
idea of a scalar field rolling down some suitable potential does not really
solve the problem because $\phi$ has to be some high energy field of a Grand
Unified Theory (GUT), and this leads to an unnaturally small value of its
mass, which is beyond all conceivable standards in Particle Physics. As an
example, utilizing the simplest form for the potential of the scalar field,
$V(\phi)=m_{\phi}^{2}\,\phi{^{2}}/2$, the present value of the associated
vacuum energy density is $\rho_{\Lambda}=\langle V(\phi) \rangle\sim
10^{-11}\,eV^{4}$, so for $\langle\phi\rangle$ of order of a typical GUT scale
near the Planck mass, $M_{P}\sim10^{19}$ GeV, the corresponding mass of $\phi$
is expected in the ballpark of $m_{\phi}\sim\,H_{0}\sim10^{-33}\,eV$.

Notice that the presence of such a tiny mass scale in scalar field models of
DE is generally expected also on the basis of structure formation
arguments~\cite{Mota04,Nunes06,Basi09}; namely from the fact that the DE
perturbations seem to play an insignificant role in structure formation for
scales well below the sound horizon. The main reason for this homogeneity of
the DE is the flatness of the potential, which is necessary to produce a
cosmic acceleration. Being the mass associated to the scalar field fluctuation
proportional to the second derivative of the potential itself, it follows that
$m_{\phi}$ will be very small and one expects that the magnitude of DE
fluctuations induced by $\phi$ should be appreciable only on length scales of
the order of the horizon. Thus, equating the spatial scale of these
fluctuations to the Compton wavelength of $\phi$ (hence to the inverse of its
mass) it follows once more that $m_{\phi}\lesssim\,H_{0}\sim10^{-33}~{\rm eV}$.

Despite the above difficulties there is a class of viable models of
quintessence usually called dilatonic
that are based on supersymmetry, supergravity and string-theory and which can
protect, for specific potentials, the light mass of quintessence (for a
review see Ref.~\cite{Ame10} and references therein). In particular,
in string theory, gauge and gravitational couplings are connected
with the vacuum expectation value of scalar field called dilaton
$\phi$~\cite{Gas}. In this context, it has been proposed by some of us
a specific model for the DE. This DE model being associated
with a rolling dilaton field that is a remnant of this non-equilibrium
phase described by a generic non-critical (Liouville)
string theory~\cite{emnw,diamandis,diamandis2,lmn}.
We call this scenario either $Q$-cosmology or
Liouville cosmology. This pattern is based on non-critical
(Liouville) strings~\cite{aben,ddk,emn}, which offer a mathematically
consistent way of incorporating time-dependent backgrounds in string theory.
We note in passing that the presence of time dependent dilaton fields at late eras of the Universe, that characterizes $Q$-cosmology, may lead to phenomenologically interesting extensions of minimal supergravity models. The latter predict less 
dark-matter relic abundances than the conventional $\Lambda$CDM model, thereby allowing for more room for supersymmetry in collider (such as LHC) tests of the models~\cite{lmn,lahanas2010}.

In this article we investigate the main dynamical properties of the
$Q$-cosmological model from the point of view of current astrophysical constraints, extending non-trivially earlier analyses~\cite{mitsou,mitsou2}.
Initially, a joint
statistical analysis, involving the latest observational data
[SNIa~\cite{Kowal08}, $H(z)$~\cite{Stern10} and BAO~\cite{Eis05,Perc10,Kazin10}] is implemented.
Secondly, we attempt to compare the matter fluctuation field of the
$Q$-cosmology with that of the
$\Lambda$CDM models by computing the growth rate of clustering.
Finally, by using the Press-Schechter formalism~\cite{press} and recently
  derived functional forms for the dark matter halo mass function~\cite{Reed},
we show that the evolution of the cluster-size halo abundances
is a potential discriminator between the $Q$-cosmology and $\Lambda$CDM
models. We would like to stress here that
the abundance of collapsed structures, as a
function of mass and redshift, 
can be accessed through observations~\cite{Evra}. Indeed, the mass
function of galaxy clusters has been measured based on X-ray surveys~\cite{Borg01, Reip02, Vik09}, via weak and strong lensing studies~\cite{Bat98, Dahle06, Corl09}, using optical surveys, like the SDSS~\cite{Bah03, Wen10}, as well as, through Sunayev-Zeldovich (SZ)
effect~\cite{Taub05}. In the last decade many authors
have found that the abundance
of collapsed structures is affected by the presence of a dark energy component~\cite{Wein03,Liberato,manera,Abramo07,Fran08,Sch09,Mort09,Rap10,Pace10,Alam10,Khed10,BPL10,Lomb10}.

The structure of our paper is as follows.
The basic elements of
the Liouville cosmological model are presented
in Section~\ref{sec:theo}, where we
also introduce the cosmological equations for a flat
Friedmann-Lemaitre-Robertson-Walker (FLRW) geometry.
In Section~\ref{sec:likelihood}, a joint statistical analysis
based on SNIa, $H(z)$ and BAO
is used to constraint the Liouville model
free parameter. The issue related with the effective DE equation
of state parameter and thus with the cosmic expansion
is presented in Section~\ref{sec:expa}.
The linear growth factor of matter perturbations
is discussed in Section~\ref{sec:growth}, while in Section~\ref{sec:cluster}, we discuss
and compare the corresponding theoretical predictions regarding the evolution of
the cluster abundances.
Finally, the main conclusions are
summarized in  Section~\ref{sec:conclu}.

\section{Theoretical Background on Q-Cosmology}\label{sec:theo}
In non-critical $Q$-cosmological models  the
(effective, four-dimensional) Friedman
equation is modified, including, apart from the standard
matter and dark energy contributions denoted
collectively by $\rho$, also Liouville-string
{\it off-shell} corrections, $\Delta \rho$,
which are not positive definite in general
(without affecting, however, the
overall positive-energy conditions of this non-equilibrium theory):
\begin{equation}
H^2(z) = \frac{8\pi G_N}{3}\rho + \Delta \rho
\label{friedmann}
\end{equation}

The detailed form of the system of dynamical equations, one of which
is (\ref{friedmann}) is given by Ref.~\cite{diamand}
in the Einstein frame~\cite{aben}, i.e.\
in the frame where the scalar curvature in the (off-shell) target space
effective action assumes the canonical Einstein form to leading order
in the Regge-slope $\alpha '$ expansion,
are given by
\bear
&&3 \; H^2 - {\tilde{\varrho}}_m - \varrho_{\phi}\;=\; \frac{e^{2
    \phi}}{2} \; \tilde{\cal{G}}_{\phi} \nonumber  \\
&&2\;\dot{H}+{\tilde{\varrho}}_m + \varrho_{\phi}+
{\tilde{p}}_m +p_{\phi}\;=\; \frac{\tilde{\cal{G}}_{ii}}{a^2} \nonumber  \\
&& \ddot{\phi}+3 H \dot{\phi}+ \frac{1}{4} \; \frac{\partial { \hat{V}}_{all} }{\partial \phi}
+ \frac{1}{2} \;( {\tilde{\varrho}}_m - 3 {\tilde{p}}_m )= \nonumber \\
&& - \frac{3}{2}\; \frac{ \;\tilde{\cal{G}}_{ii}}{ \;a^2}- \,
\frac{e^{2 \phi}}{2} \; \tilde{\cal{G}}_{\phi}  \; ,\label{eqall}
\eear
with
\bear && \tilde{\cal{G}}_{\phi} \;=\; e^{\;-2 \phi}\;( \ddot{\phi} -
{\dot{\phi}}^2 + Q e^{\phi} \dot{\phi})  \nonumber \\ &&
\tilde{\cal{G}}_{ii} \;=\; 2 \;a^2 \;\left(\; \ddot{\phi} + 3 H \dot{\phi}
+ {\dot{\phi}}^2 + ( 1 - q ) H^2 + \right. \nonumber \\ && \left. Q e^{\phi} ( \dot{\phi}+ H )\;\right) \;
,  \eear
and $q$ is the deceleration $q \equiv - \ddot{a} a /
{\dot{a}}^2$.  Above, an overdot denotes derivative with respect to
the FLRW cosmic time, $t$, $\tilde \rho_{m}$, $\tilde p_{m}$ denote
``matter'' energy and pressure densities respectively,
whilst
\bear \label{dilatonener}&& \varrho_{\phi}\;=\;\frac{1}{2}\;( \; 2
{\dot{\phi}}^2+{\hat{V}}_{all} ) \nonumber \\
&&p_{\phi}\;=\;\frac{1}{2}\;( \; 2 {\dot{\phi}}^2-{\hat{V}}_{all} ) 
\eear
denote the dilaton energy density and pressure.
In (\ref{dilatonener}), the potential has the form
$\; {\hat{V}}_{all}= 2 Q^{\;2} \exp{\;( 2 \phi )}+V \;$,
where $Q=Q(t)$ is the so-called central charge deficit
function of the non-critical theory,
which plays the role of a
``gravitational friction'' coefficient. The term $V$
represents higher-string loop corrections in the target
space-string effective action, which are in general not
known in a closed form. In our work the precise form of
$V$ will not play a role, at least for late eras of
the Universe, where such quantum string corrections are sub-dominant.
The variation of the
central charge deficit
$Q(t)$ with the cosmic time $t$ is provided by the
following consistency equation, to leading order in the Regge-slope ($\alpha '$) expansion:
\bear \frac{d \tilde{\cal{G}}_{\phi} }{d t} \;=\; - 6\; e^{\;-2
\phi}\;( H + \dot{\phi} ) \; \frac{ \;\tilde{\cal{G}}_{ii}}{ \;a^2} \;
.  \label{CXP} \eear
The reader should notice that, although we
have assumed a (spatially)
flat Universe,  the terms on the r.h.s.\ of (\ref{eqall}), which manifest
departure from criticality, act in a sense like curvature terms
as being non-zero at certain epochs.

The continuity equation for matter, which
follows by combining the set of equations (\ref{eqall}), reads:
\bear
&& \frac{d {\tilde{\varrho}}_m }{dt}+ 3 H ( {\tilde{\varrho}}_m +{\tilde p}_m) +
\frac{\dot Q}{2} \frac{\partial { \hat{V}}_{all} }{\partial Q} -
\dot{\phi}\;({\tilde{\varrho}}_m - 3 {\tilde{p}}_m )\; \nonumber \\
&& = 6\;(H+\dot{\phi})\; \frac{ \;\tilde{\cal{G}}_{ii}}{a^2} \; . \label{contin}
\eear
To proceed further one needs to make some extra assumptions.
First we assume that the matter-energy density is split as
\bear
\label{split}
\tilde{\varrho}_m = {\varrho}_d+{\varrho}_r+{\varrho}_{\delta}~.
\eear
The first term on the right-hand-side of (\ref{split}) refers to as
``dust'', $w_d=0$, and includes the baryonic matter
and any other sort of matter, characterized by $w=0$, which does not feel the
effect of the non-critical terms.
Following Ref.~\cite{diamand} we may assume that dust does not couple to
the dilaton and the non-critical-stringy corrections at late eras,
 so that its respective density,  $\varrho_{d}$, obeys the standard
 energy conservation equation  (\emph{c.f}.\ (\ref{contin})):
\begin{equation}\label{dust}
\frac{d {{\varrho}}_d }{d t}  + 3 H  {\varrho}_d \simeq 0~.
\end{equation}
The second term in (\ref{split}) refers to radiation $w_r=\frac{1}{3}$ and the
third term to an unknown sort of {\emph{exotic}} matter
which is characterized by an
equation of state (hereafter EoS)
having a weight $w_\delta$.  In our analysis for late epochs of the
Universe, such as the eras of structure formation, we may assume that
this sort of matter which would otherwise feel the non-critical terms
in the evolution of the Universe, feels predominantly only the
time-dependent dilaton $\dot \phi$  terms in (\ref{contin}),
\emph{i.e}.\ its density, $\varrho_{\delta}$, obeys approximately the
following energy conservation equation:
\begin{eqnarray}\label{newcontin}
\frac{d {{\varrho}}_\delta }{d t}  + 3 H \left(1 - \frac{\dot
    \phi}{3H} + w_\delta (z) \left( 1 + \frac{\dot \phi}{H} \right)
\right) {\varrho}_\delta \simeq 0~, \end{eqnarray}
for late eras of the Universe, \emph{e.g}.\ corresponding to redshifts
$z \le 2~,$ where analytic treatments of the Liouville $Q$-cosmology are
available. In general the exotic matter equation of state $w_\delta
(z)$ depends on the redshift, but in
our analysis in this work we consider it to be a constant.

For late epochs of the Universe, an approximate scaling behavior of the dilaton field may be found by
making the further assumption that its configuration is not affected
much by the presence of matter. In such a case, one may use the purely
gravitational solution of Ref.~\cite{diamandis}, which expresses the
dilaton in the form\footnote{A similar behavior of the dilaton is
  also valid in the early Universe as discussed in
  \cite{lahanas2010}. Such forms depend crucially on the form of the
  dilaton potential, which should be of exponential form $V \sim e^{-c
    \phi}$, $c$ some constant. Dilaton $Q$-cosmologies  are indeed
  characterized by such potentials~\cite{diamand,mitsou}.}:
\begin{equation}\label{dilaton}
\phi \simeq c_1 + \phi_0 \ln (a(t))~,
\end{equation}
in units of the present scale factor $a_0=1$, where $\phi_0, c_1$ are
constants.  In the purely gravitational theory of \cite{diamandis},
$\phi_0 = -1$, but here
we want to be more general, in order to account for the presence of matter, and so we shall consider the
constant $\phi_0$ as a phenomenological parameter to be determined by
fitting the data. However, we shall assume $\phi_0 < 0$, which assumes
a perturbatively weak string coupling $g_s=e^\phi$ as the cosmic time
elapses, which is crucial for the validity of our analytical treatment
at late epochs of the Universe. In particular, this implies that
string loop corrections, although non negligible, and in fact their
inclusion may even lead to some interesting physical effects (see
below)~\cite{mitsou}, nevertheless they can be treated
perturbatively, and thus can lead to analytic solutions at late eras.

With the above in mind, we thus observe that from (\ref{dilaton}) and (\ref{newcontin}) we obtain:
\begin{eqnarray}\label{newcontin2}
\frac{d {\varrho}_\delta }{d t}  + 3 H \left(1 - \frac{\phi_0}{3} +
  w_\delta \left( 1 + \phi_0  \right) \right) {\varrho}_\delta \simeq
0~, \end{eqnarray}
implying the following scaling behavior of the exotic matter with the scale factor:
\begin{equation}\label{defdelta}
{\varrho}_\delta \sim a^{-\delta} ~, \quad \delta =3\left(
  1-\frac{\phi_0}{3} + w_\delta \left( 1 + \phi_0  \right)\right)~.
\end{equation}
 In particular, it has been found that
$\delta \sim 4$~\cite{mitsou2} can fit the BAO data~\cite{Eis05}. Such
a value is in the ballpark of theoretically expected values.
For instance,  the simplest
phenomenological assumption on the pertinent
EoS parameter $w_\delta = \frac{p_\delta}{\varrho_\delta}$ of the
exotic dark matter fluid, consistent with
Big-Bang-nucleosynthesis constraints, is that
$w_\delta  \sim 0.4$ ~\cite{diamand}. To obtain $\delta \sim 4$, while
maintaining the feature $\phi_0 < 0$,  one must use either values of
$|\phi_0| \ll 3$, implying a weak dependence of the dilaton on the
scale factor, or $\phi_0 \sim -1$ and more generally the condition
$$ \frac{\dot \phi}{H} \sim  -1~, $$
in which case (\ref{newcontin}), (\ref{newcontin2}) become independent
of $w_\delta$.

On the other hand, the BBN constraint is also satisfied in models
where the dilaton dominates the early eras of the
Universe~\cite{lahanas2010}, in the absence of non-critical string
effects, with the exotic matter behaving as dust $w_\delta =0$ coupled
to the dilaton.  In this case,
the scaling exponent (\ref{defdelta}) becomes: $\delta \simeq 3 +
|\phi_0|$. The phenomenologically likely~\cite{mitsou2} value $\delta
\sim 4$ is then achieved for $|\phi_0| \sim 1 $, which is in the range
of the analytical solutions for late eras found in \cite{diamandis},
and for which, as already mentioned, the energy equation
(\ref{newcontin2}) is largely independent of $w_\delta$.

Following the above discussion, we therefore parametrize the normalized
Hubble parameter of this model, $E(z)=H(z)/H_{0}$, where $z$ is
the redshift, as follows~\cite{mitsou}:
\begin{eqnarray}
E(z)=\left[ {\Omega }_3 (1 + z)^3 + {\Omega }_\delta (1 +
z)^\delta + {\Omega}_2 (1 + z)^2 \right]^{1/2} 
\label{formulaforfit_txt}
\end{eqnarray}
with
\begin{equation}
\label{sumomegas_txt}
{\Omega}_3 + {\Omega}_{\delta} + {\Omega}_2 = 1.
\end{equation}
 It is interesting to mention that the current normalized Hubble function
Eq.(\ref{formulaforfit_txt}) has a form which is close to a polynomial
one. Note that a quadratic 
polynomial dark energy has been proposed by Sahni \emph{et al.} \cite{Sah02}.
We would like to stress once again
that the above
formulas are valid for late eras, such as the ones pertinent to
the supernova and other data ($0 \le z \le 2$).

In this framework, the various $\Omega_{i}$ contain
contributions from \emph{both} dark energy and matter energy
densities. Specifically, $\Omega_{3}$ does not merely represent
ordinary matter effects, but also receives contributions from the
dilaton dark energy. In fact, the sign of $\Omega_{3}$ depends on
details of the underlying theory, and it could even be \emph{negative}.
In a similar vein, the exotic contributions scaling as $(1 + z)^{\delta}$
are affected by the off-shell Liouville terms of $Q$-cosmology.
It is because of the similar \emph{scaling behaviors} of dark
matter and dilaton dark energy that we reverted to the notation
$\Omega_{i},\, i = 2,3,\delta$ in Refs.~\cite{mitsou,mitsou2}.
In order to distinguish ordinary matter from
dilaton dark energy effects that scale similarly with the
redshift, one would have to perform also measurements of the
effective dark energy EoS parameter (see Section~\ref{sec:expa}).

\section{Likelihood Analysis}\label{sec:likelihood}
Let us now discuss the statistical treatment of the
observational data used
to constrain the Liouville model presented in the previous section.
To begin with, we consider the {\em Union} supernovae Ia set of Kowalski
et al. \cite{Kowal08} which contains 307 entries.
The likelihood estimator is determined by a $\chi^{2}_{\rm SNIa}$ statistics:
\begin{equation}
\label{chi22} \chi^{2}_{\rm SNIa}=
\sum_{i=1}^{307} \left[
\frac{ {\cal \mu}_{\rm th} (z_{i},{\Omega_{3},\Omega_{\delta}})-
{\cal \mu}_{\rm
obs}(z_{i}) } {\sigma_{i}} \right]^{2},
\end{equation}
where $z_{i}$ is the
the observed redshift, ${\cal \mu}$ is the distance modulus
${\cal \mu}=m-M=5{\rm log}d_{L}+25$ and $d_{L}$ is the
luminosity distance,
\begin{equation}
d_{L}(z)=(1+z)r(z) \;\;\;\;\;\;r(z)=c\int_{0}^{z}
\frac{{\rm d}y}{H(y)}\;.
\end{equation}

In addition to the SNIa
data, we also consider the BAO scale produced in the last
scattering surface by the competition between the pressure of the
coupled baryon-photon fluid and gravity. The resulting acoustic
waves leave (in the course of the evolution) an overdensity
signature at certain length scales of the matter distribution.
Evidence of this excess was recently found in the clustering
properties of SDSS galaxies \cite{Eis05,Perc10,Kazin10}
and it provides a suitable ``standard ruler'' for constraining dark energy
models. We would like to remind the reader
that in the Liouville cosmology
the ordinary matter component which appears in the normalized Hubble parameter
Eq.~(\ref{formulaforfit_txt}) is mixed with the ``exotic'' contributions.
Thus we can not use the measurements of BAO derived by
\cite{Eis05,Perc10,Kazin10}. To overcome this
problem Mavromatos \& Mitsou \cite{mitsou2} have
proposed a different estimator which is valid for Liouville cosmologies.
This is
\be
B\equiv \left[ \left(  \int_{0}^{z_{\rm BAO}} \frac{dz}{E(z)}\right)^{2}
\frac{z_{\rm BAO}}{E(z_{\rm BAO})} \right]^{1/3}
\ee
where $z_{\rm BAO}=0.35$. Note that the measured value is
$B=0.334\pm 0.021$ \cite{mitsou2}.
Therefore, the corresponding
$\chi^{2}_{\rm BAO}$ function can be written as:
\begin{equation}
\chi^{2}_{\rm BAO}=\frac{[B(\Omega_{3},\Omega_{\delta})-0.334]^{2}}
{0.021^{2}}\;.
\end{equation}

Finally, a very interesting geometrical probe of
dark energy is provided by the
measures of $H(z)$ \cite{Stern10} from the
differential ages of passively
evolving galaxies (hereafter $H(z)$ data). Note that the sample
contains 11 entries spanning a redshift range of $0\le z<2$.
In this case the corresponding $\chi^{2}_{H}$ function can be written as:
\begin{equation}
\chi^{2}_{\rm H}=\sum_{i=1}^{11} \left[ \frac{
H_{\rm th}(z_{i},\Omega_{3},\Omega_{\delta})-H_{\rm obs}(z_{i})}
{\sigma_{i}} \right]^{2} \;\;,
\end{equation}
where $H(z)$ is the
Hubble parameter\footnote{The Hubble constant
is $H_{0}=73.8\pm 2.4$~km/s/Mpc~\cite{Riess10}.}, $H(z)=H_{0}E(z)$.

In order to put tighter constraints on the corresponding
parameter space of our cosmological model, the above probes are combined
through a joint likelihood analysis\footnote{Likelihoods are
normalized to their maximum values. In the present analysis we
always report $1\sigma$ uncertainties on the fitted parameters. Note
also that the total number of data points used here is
$N_{\rm tot}=319$, while the associated degrees of freedom are: ${\text{\em
  dof}} = 317$. Note that we sample
$\Omega_{3} \in [-4,1]$ and
$\Omega_{\delta} \in [0.1,2]$ in steps of
0.001.}, given by the product of the individual likelihoods
according to: ${\cal L}_{\rm tot}= {\cal L}_{\rm SNIa}\times
{\cal L}_{\rm BAO} \times {\cal L}_{\rm H}$, which translates in the
joint $\chi^2$ function in an addition:
$\chi^{2}_{\rm tot}=\chi^{2}_{\rm SNIa}+\chi^{2}_{\rm
BAO}+\chi^{2}_{\rm H}$. The resulting best fit parameters for
different values of
$\delta$, are presented in Table~\ref{tab:chi}. The current statistical results are
in very good agreement with those found by \cite{mitsou2}.

\begin{table}[ht]
\caption[]{Results of the likelihood function analysis.
The $1^{\rm st}$ column indicates the Liouville model.
$2^{\rm nd}$ column presents the values of $\delta$ used.
$3^{\rm rd}$, $4^{\rm th}$ and $5^{\rm th}$
columns show the best fit parameters and the reduced
$\chi^{2}_{\rm tot}$. In the final column one can find various
line types of the models appearing in Figs.~\ref{fig:hubble} and~\ref{fig:expa}.}\tabcolsep 3.pt
\vspace{1mm}
\begin{tabular}{cccccc} \hline \hline
Model & $\delta$ & $\Omega_{3}$& $\Omega_{\delta}$& $\chi^{2}_{\rm tot}/317$&Symbols \\ \hline
$Q_{1}$ & $4.3$ & $-1.43\pm 0.14$ &$0.27\pm 0.02$ &1.030& solid\\
$Q_{2}$ & $4.1$ & $-1.64\pm 0.16$ &$0.40\pm 0.04$ &1.028& dashed \\
$Q_{3}$ & $3.9$ & $-1.93\pm 0.19$ &$0.59\pm 0.06$ &1.027& dotted \\
$Q_{4}$ & $3.7$ & $-2.45\pm 0.25$ &$0.95\pm 0.10$ &1.025& triangles\\
$Q_{5}$ & $3.5$ & $-3.29\pm 0.33$ &$1.63\pm 0.17$ &1.024& open cirles \\ \hline\hline \label{tab:chi}
\end{tabular}
\end{table}

The reader may worry that negative dust-like energy densities indicate
an instability of the model and an obvious violation of the positive
energy conditions, and thus its immediate exclusion. However,
as discussed in \cite{mitsou,mitsou2}, and already mentioned in the
previous section, the ``dust''-like contributions, $\Omega_3$, do not merely represent
ordinary matter effects, but also receive contributions from the
dilaton dark energy. In fact, the sign of $\Omega_3$ depends on
details of the underlying theory, and it could even be negative. For
instance, as argued in \cite{mitsou},
string loop corrections could lead to a negative $\Omega_3$.
In addition,
Kaluza-Klein graviton modes in certain brane-inspired
models~\cite{kaluza} also yield negative dust contributions. 
In a similar vein, the exotic contributions scaling 
as $(1+z)^\delta$ are affected
by the off-shell Liouville terms of $Q$-cosmology.

\section{The recent expansion history}\label{sec:expa}
\begin{figure}[ht]
\includegraphics[width=0.5\textwidth]{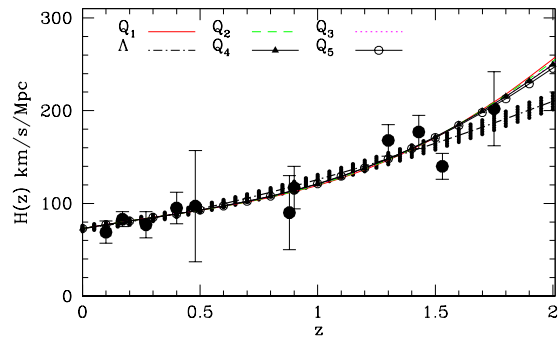}
\caption{Comparison of the observed (solid circles~\cite{Stern10})
and theoretical evolution of the Hubble parameter,
$H(z)$, using $H_{0}=73.8\pm 2.4$~km/s/Mpc~\cite{Riess10}.
The different Liouville models are
represented by different symbols and line types (see Table~\ref{tab:chi}
for definitions).
For comparison, the dot-dashed line corresponds to
the traditional $\Lambda$CDM model, while the thick-line error bars 
correspond to $1\sigma$ $H_{0}$-uncertainties.
We do not plot the $1\sigma$ $H_{0}$-uncertainties for the Liouville models 
in order to avoid confusion.
}\label{fig:hubble}
\end{figure}

In Fig.~\ref{fig:hubble} we plot the Hubble parameter of the current Liouville models
as a function of redshift, which for redshifts $z \gtrsim 1.5$ appears
to be different in amplitude
with respect to the corresponding $\Lambda$CDM model expectations
(dot-dashed line).

As we have already mentioned in Section~\ref{sec:theo}
the density related with the $\Omega_{3}$ parameter
varies as $\varrho_{3}(z) \propto (1+z)^{3}$.
Owing to the fact that the density of
ordinary matter (baryons and usual cold dark matter), denoted $\varrho_m$
from now on, follows the same power-law
as $\varrho_{3}(z)$, $\varrho_{m}(z) \propto (1+z)^{3}$,
we are completely justified to assume that
the usual matter density is one of the components
of the density $\varrho_{3}(z)$.
In other words the
$\Omega_{m}(z)=\varrho_{m}(z)/\varrho_{c}(z)$
density parameter (where $\varrho_{c}(z)=3H^{2}(z)/8\pi G$
is the critical density parameter) is
included in the $\Omega_{3}(z)$ parameter.
On the other hand, as a result of the existence of non-trivial dilaton
couplings to some dark matter species, as mentioned previously,
there is dark matter with exotic scaling $\Omega_\delta (1 +
z)^\delta$, with $\delta$ given by (\ref{defdelta}).
Although this is \emph{not} a dark energy contribution, nevertheless it is the dilaton
couplings that modify its scaling, and therefore in this sense, the
origin of this exotic scaling may be attributed to the existence of
the interaction of dilaton dark energy with dark matter. However,
since the $\Omega_\delta > 0$ represents massive matter,
it may cluster and lead to non trivial contributions to structure
growth in the Universe. In this sense,  it is distinguishable from
dark energy contributions that are non clustering. One of our points
in this work is an attempt to distinguish
phenomenologically whether the current astrophysical
data point out towards clustering of this exotic
matter, and hence contribution to structure formation and growth, or not.

To this end, we first write the normalized
Hubble parameter Eq.~(\ref{formulaforfit_txt}) as follows
\begin{equation}\label{parHubMass}
E(z)=\frac{H(z)}{H_{0}}=\left[\Omega_{m}(1+z)^{3} +  \Delta H(z)^2 \right]^{1/2}
\end{equation}
with
 \begin{equation}\label{parHubMass1}
\Delta H(z)^2 = \Omega_{\delta}(1+z)^{\delta}+ (\Omega_{3}-\Omega_{m})(1+z)^{3}+
\Omega_{2}(1+z)^{2}.
 \end{equation}
 Notice here that in the above formula
 we have been careful to isolate the matter contributions, pertaining to
dust $\Omega_d$ (for baryons and dust dark matter
$\Omega_i=\varrho_{i0}/\varrho_{c0}$), from  the exotic matter terms,
$\Omega_\delta $ with non-trivial equation of state $w_\delta$,
corresponding to a scaling exponent $\delta$ (\ref{defdelta}).
It should be stressed that the last term $\Delta H^2(z)$ of the
normalized Hubble function (\ref{parHubMass}) encode
the $Q$-cosmology corrections to the standard FLRW expression.

Since the corresponding components exhibit smooth scaling with the redshift,
the absorption of these extra effects to an ``effective dark energy''
contribution, with a non-trivial EoS is possible.
In general, it is well known that one can express
the effective dark energy EoS parameter in terms of the
normalized Hubble parameter \cite{Saini00}
\begin{equation}
\label{eos22}
w_{\rm DE}(a)=\frac{-1-\frac{2}{3}\frac{d\ln E}{d\ln a}}
{1-\Omega_{m}(a)}\;,
\end{equation}
where in our case
\be
\label{omm}
\Omega_{m}(a)=\frac{\Omega_{m}a^{-3}}{E^{2}(a)} \;.
\ee

Differentiating the latter
and utilizing
Eq.~(\ref{eos22}) we find that
\be
\label{domm}
\frac{d\Omega_{m}}{d\ln a}=
3w_{\rm DE}(a)\Omega_{m}(a)\left[1-\Omega_{m}(a)\right]\;.
\ee
Notice that we will use the above quantity in Section~\ref{sec:growth}.

After some simple algebra, it is also readily seen that
the effective dark energy EoS parameter is
given by (see \cite{Linjen03}):
\begin{equation}
\label{eos222}
w_{\rm DE}(a)=-1-\frac{1}{3}\;\frac{d\ln \Delta
H^{2}}{d\ln a}
\end{equation}
or
\begin{equation}
\label{eos223}
w_{\rm DE}(z)=-1+\frac{1}{3}\;\frac{d\ln \Delta
H^{2}}{d\ln (1+z)}
\end{equation}
where $a=(1+z)^{-1}$. 
From the above analysis it becomes 
clear that any modifications to the EoS parameter are included in the second 
term of Eq.~(\ref{eos223}). A similar analysis for other 
DE models can be found in \cite{Linjen03,Linder05}.

In our case, on inserting Eq.~(\ref{parHubMass1}) into
Eq.~(\ref{eos223}) it is straightforward to obtain a simple analytical
expression for the effective dark energy EoS parameter:
{\small \begin{equation}\label{eos221}
w_{\rm DE}(z) = -  1 + \frac{1}{3} \;
\frac{3(\Omega_{3}-\Omega_{m})(1+z)+\delta \Omega_{\delta}(1+z)^{\delta-2}
+2\Omega_{2}}
{(\Omega_{3}-\Omega_{m})(1+z)+\Omega_{\delta}(1+z)^{\delta-2}
+\Omega_{2}} \;.
\end{equation}}

It thus follows that in the current cosmological context, the Liouville
scenario as proposed by Ellis \emph{et al}.~\cite{mitsou} can be
treated by an additional effective fluid with EoS parameter
defined by Eq.~(\ref{eos221}).
In order to investigate the behavior of the
effective EoS parameter we have to know a priori
the value of $\Omega_{m}$. The matter density $\Omega_m$
remains the most weakly constrained cosmological parameter.
In principle, $\Omega_m$ is constrained by the maximum
likelihood fit to the WMAP~\cite{komatsu08} and SNIa~\cite{Kowal08} data in the context of the
concordance $\Lambda$CDM cosmology. However, in the
spirit of this work, we want to use measures which are completely
independent of the dark energy component. An estimate of $\Omega_m$
without conventional priors is not an easy task in observational cosmology.
Nevertheless, various authors, using mainly large scale structure studies,
have attempted to set constraints on the $\Omega_m$ parameter.
In a rather old paper, Plionis \emph{et al}. \cite{PLC}, using the motion
of the Local Group with respect to the cosmic microwave background,
found that $\Omega_{m}\simeq 0.30$. 
From the analysis of the power spectrum, Sanchez \emph{et al}. \cite{San06}
obtained a value
$\Omega_m\simeq 0.24$. Moreover, the authors of Refs.~\cite{Fel03,MoTu05} analyzed the peculiar velocity
field in the local Universe and obtained the values
$\Omega_m\simeq 0.30$ and $\simeq 0.22$ respectively.
In addition, the authors of Ref.~\cite{And05},
based on the cluster mass-to-light ratio,
claim that $\Omega_m$ lies in the interval $0.15-0.26$ (see
also \cite{Sch02} for a review).
Therefore, there are strong independent indications that $\Omega_m$ lies in the range  $0.2\lesssim
\Omega_m\lesssim 0.3$.  In order to compare our results with
those of the flat $\Lambda$CDM model, we shall restrict our present analysis
to an indicative value of $\Omega_{m}=0.28$.

\begin{figure}[ht]
\includegraphics[width=0.5\textwidth]{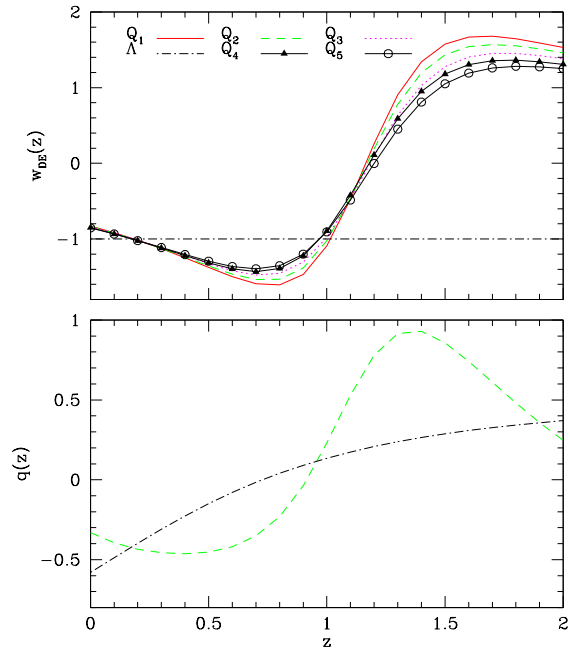}
\caption{Expansion history. In the upper panel
we display the evolution of the dark energy effective
EoS parameter. The different Liouville models are
represented by different symbols and line types (see Table~\ref{tab:chi}
for definitions). In the lower panel we compare the
deceleration parameters of the $Q_{2}$
and the concordance $\Lambda$CDM (dot-dashed line) models.
The deceleration parameter is plotted only for the $Q_{2}$
model in order to avoid confusion.}\label{fig:expa}
\end{figure}

In order to visualize the redshift dependence of the
effective EoS parameter, we compare in the upper panel of Fig.~\ref{fig:expa}
various flat Liouville cosmological models (see Table~\ref{tab:chi}).
One can divide the evolution of the cosmic expansion history in different
phases on the basis of the varying behavior of the
Liouville and $\Lambda$CDM models. We will investigate such variations
in terms of the deceleration parameter, $q(z)=-[1-d\ln H/d\ln(1+z)]$, which is plotted in the lower panel of Fig.~\ref{fig:expa}.
We can divide the cosmic expansion history in the following phases:
\begin{itemize}
\item At early enough times $z\sim 2$ the deceleration
      parameters of both models are positive with $q_{Q}<
      q_{\Lambda}$, which means that the $\Lambda$CDM model is more decelerating
      than the Liouville model.
\item For $1.6\le z<1.9$ the deceleration parameters are
    both positive with $q_{Q}>q_{\Lambda}$, which means that the
    cosmic expansion in the Liouville model is more rapidly
    ``decelerating'' than in the $\Lambda$CDM case.
    Note that the effective EoS parameter of the $Q$-model
    tends to its maximum value, $w_{\rm DE}(z) \sim 1.4$,
    while we always have $w_{\rm DE} = -1$ for
    the $\Lambda$CDM model (dot-dashed line);
\item Between $1.1\le z<1.6$ the deceleration parameters remain
      positive (with $q_{Q}<q_{\Lambda}$) but $q_{Q}$ decreases rapidly.
\item For $0.7<z<1.1$ the traditional $\Lambda$CDM model
    remains in the decelerated regime ($q_{\Lambda}>0$) but
    the Liouville model is starting to accelerate ($q_{Q}<0$, $w_{\rm DE}<-1/3$).
    At $z\simeq 0.9$ the corresponding effective EoS
    parameter cross, for the first time,
    the phantom divide ($w_{\rm DE}=-1$) and it stays there
    for some time ($0.2 <z <0.9$).
\item Prior to the present epoch the
      effective EoS parameter crosses the phantom divide
      for the second time ($z\simeq 0.2$) and then it remains close to
      $w_{\rm DE}\simeq -0.90$. The deceleration parameters are
      both negative and since $q_{Q}>q_{\Lambda}$, the
      $\Lambda$CDM model provides a stronger acceleration than in the
      Liouville model. At the present time we find $q_{Q,0}\simeq -0.33$
      and $q_{\Lambda,0}\simeq -0.60$
\end{itemize}

\section{The growth factor and the rate of clustering}\label{sec:growth}
In this section, we discuss
the basic equation which governs the behavior of the matter
perturbations on sub-horizon scales and within the framework of any
dark energy model,
including those of modified gravity (``geometrical dark energy'').

The basic assumption underlying the approach is that whatever we call
``dark energy '' contribution in our effective approach, pertains
to a substance that is non clustering to participate significantly
in the growth of cosmic structures. In our Liouville cosmology, this
issue is subtle. As we have discussed above, we do have two kinds of
matter in the model, with two different equations of state and
scaling, an ordinary dust, and an exotic dark matter
component, with non trivial
scaling exponent $\delta$ (\ref{defdelta}), as a result of its
coupling to the dilaton.
The phenomenologically likely value $\delta \sim 4$ makes the
behavior of exotic dark matter resembling (to a good approximation)
that of relativistic matter (radiation), which does not cluster.
We shall come back to this point later one. For the moment let us
restrict ourselves to the ordinary matter as the dominant kind
that contributes to structure formation via its clustering properties.

For these cases, a full analytical
description can be introduced by considering an extended Poisson equation
together with the Euler and continuity equations.
Consequently,  we consider here the evolution equation
of the matter fluctuations for models where the DE
fluid has a vanishing anisotropic stress and it is not
coupled to ordinary matter (noninteracting DE see \cite{Lue04,Linder05,Stab06,Uzan07,Linder2007,Tsu08,Gann09}).
It is well known that for small scales (smaller than the horizon)
the dark energy component (or ``geometrical'' dark energy)
is expected to be smooth and thus it is fair to consider perturbations only on the usual matter component of the
cosmic fluid \cite{Dave02}. This assumption leads to the traditional 
equation for matter perturbations (see Appendix~\ref{app:perturb}):
\be
\label{odedelta}
\ddot{\delta}_{m}+ 2H\dot{\delta}_{m} =
4 \pi G \rho_{m} \delta_{m},
\ee
where $\rho_{m}$ is the matter density which clusters in order to form
cosmic structures.

We would like to emphasize here that Eq.(\ref{odedelta}) can be used 
also in the case of a dilaton field, as it has been shown by  
Boisseau \emph{et al.} \cite{Boi00}~\footnote{The authors of \cite{Boi00} have worked in the so-called Jordan frame, in which
the dilaton couples to the Einstein sacalar curvature term in the (four space-time dimensional) action through a function $F(\phi)$: 
$$ S = \int d^4 x \frac{F(\phi)}{2} R - \frac{1}{2} g^{\mu\nu} \partial_\mu \phi \partial_\nu \phi - V(\phi) + \mathcal{L}_{\rm dust~matter}(g_{\mu\nu}) ~. 
 $$ In this formalism there is an effective gravitational ``constant'', $G_{\rm eff}(\phi)$,  which is a function of the dilaton that replaces $G$ in 
 (\ref{odedelta}).  Notice that the matter action does not couple to the dilaton in the Jordan frame. In our Liouville string cosmology case we have an additional component to the matter action, $\mathcal{L}_{\rm exotic~matter}(g_{\mu\nu}, \phi)$, involving the exotic dark matter, with equation of state $w_\delta$, and its interactions with the ordinary baryonic or dark matter with dust scaling. This depends on the dilaton in the Jordan frame. In the Liouville Cosmology we work in the Einstein frame, as already mentioned, by having redefined appropriately the gravitational field, in such a way that, in terms of the new metric field, there is a canonically normalized Einstein term
 in the effective action. Such a redefinition leaves the perturbative scattering amplitudes of low-energy string theory invariant, and hence is an allowed transformation in the string framework that does not affect the physical predictions. Under this redefinition, the matter action involves coupling of the dilaton with the dust-like matter fields, as a consequence of the redefined gravition. Nevertheless,  we still assume for our purposes that such a coupling is very weak, and the only dominant couplings of the dilaton to matter fields in the Einstein frame are the ones pertaining to the exotic dark matter components. As we shall discuss below, such assumptions can be put directly to experimental tests by studying galactic growth data.}
In other words, Eq. (\ref{odedelta}) is valid also in the Liouville cosmology, as far as ordinary dust matter (including potential dark matter components scaling like dust) is concerned.

Within this latter framework, we assume
that for small scales (smaller than the horizon)
the effective dark energy component $\Delta H(z)$ appeared in the
modified Friedmann equation (\ref{parHubMass})
is smooth (homogeneous)
which implies that Eq.~(\ref{odedelta}) is valid also in the Liouville model.
If we change the variables from $t$ to $\ln a$
($\frac{d}{dt}=H\frac{d}{d\ln a}$)
then the time evolution
of the mass density contrast (see Eq.~(\ref{odedelta})) takes the
following form~\cite{Wang98}
\be
\label{dela}
\left( \ln \delta_{m}\right)^{''}+\left( \ln \delta_{m}\right)^{'2}+
\left( \ln \delta_{m}\right)^{'}X(a)=\frac{3}{2}\Omega_{m}(a),
\ee
where
\be
X(a)=\frac{1}{2}-\frac{3}{2}w_{\rm DE}(a)[1-\Omega_{m}(a)] \;.
\ee
The prime denotes derivatives with respect to $\ln a$.
Now, for any type of dark energy
an efficient parametrization
of the matter perturbations
is based on the growth rate of clustering
\cite{Peeb93}
\be
\label{fzz221}
f(a)=\frac{d\ln \delta_{m}}{d\ln a}=\Omega^{\gamma}_{m}(a)
\ee
$\gamma$ is the so called growth index \cite{Wang98,Linjen03,Lue04,Linder2007,Nes08}
which plays 
a key role in cosmological studies, especially in light of 
recent large redshift surveys, such as the {\em WiggleZ} \cite{Blake,Sam11}. 
Indeed, it has been proposed that measurement of 
the growth index can provide an efficient
way to discriminate between modified gravity models and DE models
which adhere to general relativity. As an example, it was theoretically
shown that for DE models
within the general relativity framework, the growth index $\gamma$ is well fitted by
$\gamma_{\rm GR}\approx 6/11$ \cite{Linder2007,Nes08}.
Notice, that in the case of the
braneworld model of Ref.~\cite{DGP}
we have $\gamma \approx 11/16$ (see also \cite{Linder2007}), while for
the $f(R)$ gravity models we have $\gamma(z)\approx 0.41$ at $z=0$ 
\cite{Gann09}.

Now, inserting Eq.~(\ref{fzz221}) into Eq.~(\ref{dela}) and using simultaneously
Eq.~(\ref{domm}) we obtain (see also~\cite{Wang98})
\be
\label{fzz222}
3w_{\rm DE}(a)\Omega_{m}(a)[1-\Omega_{m}(a)]\frac{df}{d\Omega_{m}}+f^{2}+fX(a)  
= \frac{3}{2}\Omega_{m}(a)
\ee
In this framework, Nesseris \& Perivolaropoulos \cite{Nes08} used for the 
first time growth-rate data (most of the data 
are included in our Table~\ref{tab:growth}) in order to 
constraint $\gamma$. They found (see also \cite{Wang98,Linder2007}) that a good
approximation of the growth index, especially at $z\sim 1$, is given by:
\be
\label{index1}
\gamma\simeq \frac{3(w_{\rm DE}-1)}{6w_{\rm DE}-5} \;.
\ee
Of course the above formula boils down to $6/11$ for the usual
$\Lambda$CDM [$w_{\rm DE}(z)=-1$] model as it should.\footnote{We should point out at this stage that in order to compare the observed 
growth history of the Universe with predictions coming from modified gravity models, which however is not our case, one cannot use the observed $f_{obs}(z)$ but rather 
a combination of the growth rate of structure and the 
rms fluctuations of the linear density 
field on scales of 8$h^{-1}$Mpc, namely $f(z)\sigma_{8}(z)$. This is a 
model independent quantity
(see e.g.\ \cite{Sam11} and references therein). }

Since the effective EoS parameter of the Liouville cosmological model
varies strongly with redshift (see Fig.~\ref{fig:expa}) one would expect that
the corresponding growth index would be a function of redshift.
Following the notations of Polarski \& Gannouji \cite{Pol} we consider
a first order expansion of $\gamma$ in redshift
\be
\label{ind1}
\gamma(z)=\gamma_{0}+\gamma_{1}z,
\ee
where $\gamma_{0}$ is given by Eq.~(\ref{index1}). Using
Eq.~(\ref{ind1}) in (\ref{fzz221}) and (\ref{fzz222})
we obtain (see also~\cite{Pol})
\be
\gamma_{1}=\frac{\Omega_{m}^{\gamma_{0}}+3(\gamma_{0}-\frac{1}{2})w_{\rm DE, 0}
(1-\Omega_{m})-\frac{3}{2}\Omega_{m}^{1-\gamma_{0}}+\frac{1}{2}  }
{\ln  \Omega_{m}},
\ee
where $\gamma_{0}$ is given by Eq.~(\ref{index1}).
We would like to remind the reader that in the current article we use
$w_{\rm DE,0}\simeq -0.90$ and
$\Omega_{m}\simeq 0.28$
which imply that $(\gamma_{0},\gamma_{1})\simeq (0.550,-0.048)$.
We also find that the latter set of values remains quite robust
within the range of the physically accepted values of $\Omega_{m}$.
Indeed as an example in the case of $\Omega_{m}=0.30$ we find
$(\gamma_{0},\gamma_{1})\simeq (0.549,-0.046)$.

\begin{figure}[ht]
\includegraphics[width=0.5\textwidth]{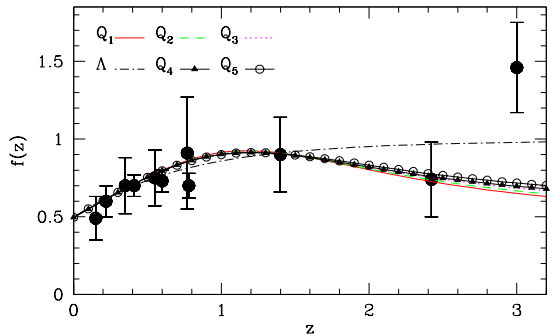}
\caption{Comparison of the observed, 
[solid circles; see Table~\ref{tab:growth}], and 
theoretical evolution of the growth rate of clustering
$f(z)$. Note that the different Liouville models are
represented by different symbols and line types (see Table~\ref{tab:chi}
for definitions). The dashed line corresponds to the 
$\Lambda$CDM cosmology.}\label{fig:growth}
\end{figure}

In Table~\ref{tab:growth}, we
quote the observational values of the growth rate of clustering, with the
corresponding error bars. This is an enriched version of the dataset 
used in \cite{Nes08},  which includes in addition data from the 
{\em WiggleZ}~\cite{Blake}. 
The observed growth rate of clustering is given by $f_{obs}=\beta b$, 
where $\beta(z)$ is the redshift space distortion 
parameter and $b(z)$ is the linear bias. Observationally, one can estimate 
the $\beta(z)$ parameter by using 
the anisotropy of the correlation function. The linear bias factor can be defined 
as the ratio of the variances of the tracer (galaxies, QSOs, etc) 
and underlying mass
density fields, smoothed out at the scale $8h^{-1}$ Mpc, at which the variance is of
order unity: $b(z)=\sigma_{8,tr}(z)/\sigma_{8}(z)$, where
$\sigma_{8,tr}(z)$ is measured directly from the sample. As discussed 
in~\cite{Nes08}, the weak point of 
the $f_{obs}(z)$ data is 
the fact that the $\sigma_{8}(z)$ is 
defined at each redshift using a particular reference (fiducial) $\Lambda$CDM 
model (see third column in Table~\ref{tab:growth}). Moreover, the growth measurements 
are correlated with the shape of the power spectrum, 
which makes the growth rate sample rather heterogeneous~\cite{Sam11,Simp10}. 
In our analysis, we ignore the effects of such a heterogeneity. 
Such simplifying assumptions in treating the growth rate data have been used extensively 
in the literature in order to put constraints on the growth 
index $\gamma$, especially in the context of the braneworld cosmological 
model~\cite{Wei}. We would like to stress that 
in the present work, due to the above 
caveats, we do not use the $f_{obs}(z)$ data to constrain 
the cosmological parameters or the growth index. Instead, 
by using the constraints found from the cosmic expansion data (see section~\ref{sec:likelihood}),
we shall only compare the evolution of the predicted Liouville
growth rate of structure with the data (for a similar analysis 
see \cite{Zhang}).

In Fig.~\ref{fig:growth} we display the
predicted growth rate Eq.~(\ref{fzz221}) together with the
observed $f_{obs}(z)$~\cite{Bas11}.
We compare the growth rate of clustering
between data and models via a $\chi^{2}$ statistical test. For the model
predictions we use the best fitted values for the parameters
($\Omega_{3}$ and $\Omega_{\delta}$) obtained in Section~\ref{sec:theo}
for each Liouville model. We then
compute the corresponding consistency between models and data
($\chi^{2}_{\rm min}/11$)
and place these results in
Table~\ref{tab:growth-chi}. For the Liouville models we find
$\chi^{2}_{\rm min}/11 \simeq 1.19$ while
in the case of the concordance $\Lambda$CDM cosmology we find
$\chi^{2}_{\rm min}/11 \simeq 0.60$.
Note that in the last column of Table~\ref{tab:growth-chi}, we list
the corresponding
results by excluding from the statistical analysis the observed
growth rate of clustering at $z=3$ \cite{McDon05}.
If we consider $\Omega_{m}=0.30$, we then find either
$\chi^{2}_{\rm min}/11 \simeq 1.47$ or $\chi^{2}_{\rm min}(z<3)/10 \simeq 0.70$.

\begin{table}[ht]
\caption[]{Data of the growth rate of clustering. The correspondence
of the columns is as follows: redshift, observed growth rate, cosmological parameters 
use by different authors and references.} \tabcolsep 4.5pt
\vspace{1mm}
\begin{tabular}{cccc} \hline \hline
z& $f_{obs,Ref}$ & $(\Omega_{m,Ref},\sigma_{8,Ref})$& Refs. \\ \hline
0.15 & $0.49\pm 0.14$&$(0.30,0.90)$ &\cite{Guzzo08,Verde02,Hawk03}\\
0.22 & $0.60\pm 0.10$& $(0.27,0.80)$&\cite{Blake}\\
0.35 & $0.70\pm 0.18$& $(0.24,0.76)$&\cite{Teg06} \\
0.41 & $0.70\pm 0.07$& $(0.27,0.80)$&\cite{Blake}\\
0.55 & $0.75\pm 0.18$& $(0.30,1.00)$&\cite{Ross07}\\
0.60 & $0.73\pm 0.07$& $(0.27,0.80)$&\cite{Blake}\\
0.77 & $0.91\pm 0.36$& $(0.27,0.78)$&\cite{Guzzo08}\\
0.78 & $0.70\pm 0.08$& $(0.27,0.80)$&\cite{Blake}\\
1.40 & $0.90\pm 0.24$& $(0.25,0.84)$&\cite{daAng08}\\
2.42 & $0.74\pm 0.24$& $(0.26,0.93)$&\cite{Viel04}\\
3.00 & $1.46\pm 0.29$& $(0.30,0.85)$&\cite{McDon05}\\ \hline\hline \label{tab:growth}
\end{tabular}
\end{table}

\begin{table}[ht]
\caption[]{Statistical quantification of the comparison between 
the observed growth rate of clustering (see Table~\ref{tab:growth})  
and the theoretical model expectations. The last column corresponds 
to comparison excluding the observed
growth rate of clustering at $z=3$ \cite{McDon05}.}
\vspace{1mm}
\tabcolsep 2.5pt
\begin{tabular}{ccc} \hline \hline
Model& $\chi^{2}_{\rm min}/11$ & $\chi^{2}_{\rm min}(z<3)/10$  \\ \hline
$Q$-Data & 1.19 &0.90\\
$\Lambda$-Data & 0.60 &0.45\\ \hline\hline \label{tab:growth-chi}
\end{tabular}
\end{table}

Finally, let us now examine the case in which
the density
$\varrho_{\delta}(z) \propto (1+z)^{\delta}$ participates in the
clustering.
The situation is subtle. As we have discussed above, the exotic
matter obeys a modified energy equation, as a result of its direct
coupling to the
dilaton sources in the conservation Eqs.~(\ref{contin}) and~(\ref{newcontin}).
Thus, in addition to the existence of pressure in the exotic
dark matter fluid, due to a possibly non-trivial equation of state, $w_\delta$,
one has the effects of the dilaton coupling that also contribute
to the growth equation.  For $\phi_0 \sim -1$ in (\ref{dilaton}),
or more generally $\dot \phi \sim -H$, the modified conservation
equation (\ref{newcontin2}) becomes independent of
the equation of state $w_\delta$:
\be\label{resrad}
\dot{\varrho}_{\delta} + 3(1+\nu)H \varrho_{\delta} \simeq 0,
\ee
where $\nu=1/3$.
This energy equation resembles that of a radiation fluid, with
effective scaling law for the density $\varrho_\delta \sim a^{-4}$.
However, the reader must bear in mind that the energy equation
(\ref{resrad}) stems from the time-dependent dilaton-sources modified
energy equation (\ref{newcontin}), which for the case at hand, $\dot
\phi \sim -H$, is independent of the true equation of state
$w_\delta$.
Within the standard Cosmological framework, the growth equation
for radiation is modified, as compared to that for matter
(\ref{odedelta}),  by terms larger by a factor of about
$(1+\nu)(1+3\nu)=8/3$~\cite{peacock}:
\be\label{radiation}
\ddot \delta_\delta  + 2 H {\dot \delta}_\delta
\simeq 4\pi G(1+\nu)(1+3\nu) \rho_\delta \delta_\delta ~.
\ee
In this case the growing mode scales as $\delta_{\delta} \sim a^{1+3\nu}$
at large redshifts which implies that the growth rate
takes a large value $f_{\delta}\sim 1+3\nu=2$.
The latter value seems to be ruled out by
the observed growth rate of clustering.

\section{Comparison with Cluster Halo abundances}\label{sec:cluster}
Since the Liouville cosmology appears not to be ruled out by the
growth rate data,
it is important to define observational criteria that will enable us
to distinguish between it and the concordance $\Lambda$CDM
cosmology. An obvious choice, that has been extensively used so far, is to
extract the theoretical predictions of the models for 
the cluster-size halo redshift distributions and to confront them 
with the data. This will allow
distinguishing the $Q$-cosmology from the $\Lambda$CDM model. 
Recently, the halo abundances predicted by a
large variety of DE models have been compared with those corresponding
to the $\Lambda$CDM model~\cite{BPL10}. As a result, it is suggested
that many DE models (including some of modified gravity)
are clearly distinguishable from the $\Lambda$CDM cosmology.

We use the Press and Schechter~\cite{press} formalism,
which determines the fraction of matter that has formed bounded
structures as a function of redshift. Mathematical details of such a
treatment can be found also in~\cite{BPL10}; here we only present the basic
ideas.
The number density of halos, $n(M,z)$,
with masses within the range $(M,\:M+\delta M)$ is given by:
\begin{equation}
\label{MF}
n(M,z) dM = \frac{\bar{\rho}}{M} \frac{d{\rm \ln}\sigma^{-1}}{dM} f(\sigma) dM \;,
\end{equation}
where $\bar\rho$ is the mean mass density. In the original Press-Schechter (PSc) formalism,
$f(\sigma)=f_{\rm PSc}(\sigma)=\sqrt{2/\pi} (\delta_c/\sigma)
\exp(-\delta_c^2/2\sigma^2)$, $\delta_{c}$ is the
linearly extrapolated density
threshold above which structures collapse~\cite{eke}, while
$\sigma^2(M,z)$ is the mass variance of
the smoothed linear density field, extrapolated to redshift $z$ at which
the halos are identified. It depends on the power-spectrum of
density perturbations in Fourier space, $P(k)$, for which we use here the
CDM form according to \cite{Bard86}, and the values of the
baryon density parameter, the spectral slope and Hubble constant
according to the recent WMAP7 results \cite{komatsu08}.
Although the original Press-Schechter mass-function, $f_{\rm PSc}$,
was shown to provide a good first approximation to that
provided by numerical simulations, it was later found
to over-predict/under-predict the number of low/high mass halos at the
present epoch \cite{Jenk01,LM07}.
More recently,  a large number of works have provided better fitting
functions of $f(\sigma)$, some of them  based on a phenomenological
approach. In the present
treatment, we adopt the one proposed by Reed \emph{et al}. \cite{Reed}.

We remind the reader that it is customary to parametrize the mass
variance in terms of $\sigma_8$, the rms mass fluctuations on scales of 
$8h^{-1}$ Mpc at redshift $z=0$. In order to compare the mass function predictions of the
different cosmological models, it is imperative to use
for each model the corresponding value of $\sigma_8$ and $\delta_c$.
It is well known that for the usual $\Lambda$CDM cosmology
$\delta_{c} \simeq 1.675$,
while Weinberg \& Kamionkowski \cite{Wein03} provide an accurate fitting
formula to estimate $\delta_{c}$ for any DE model
with a constant equation of state parameter (see their Eq.18).
Although these conditions are not satisfied by our models,
one can show~\cite{Pace10,BPL10} that the $\delta_c$ values
for a large family of dark energy models with a time-varying EoS
parameter can be well approximated using the previously discussed
fitting formula, as long as the EoS parameter remains
constant near the present epoch.
Since for the current Liouville cosmological model ($Q_2$ with $\Omega_m=0.28$)
the effective dark energy EoS parameter remains
close to $w_{\rm DE}\simeq -0.90$, prior to the present time, implies
that we can use the approximate
Weinberg \& Kamionkowski \cite{Wein03} fitting formula.
Doing so we find
$\delta_{c} \simeq 1.668$. As a further consistency check we obtain
the $\delta_{c}$ value using the same methodology with
that of Pace \textit{et al.}\ (\cite{Pace10}; see their Eq.~(18)).
We find $\delta_{c} \simeq 1.676$, which is in relatively
good agreement with that value found by the fitting
formula of \cite{Wein03}. Note that for $\Omega_{m}=0.3$ we obtain
$\delta_{c} \simeq 1.674$. Below we utilize those $\delta_{c}$
values which are based on the notation of Pace \emph{et al}. \cite{Pace10}.

Now, in order to estimate the correct normalization of the model's
$\sigma_8$ power spectrum, 
we use the formulation presented in Ref.~\cite{BPL10} which scales the
observationally determined $\sigma_{8, \Lambda}$ value to that of any
cosmological model. The corresponding value for the Liouville model is
$\sigma_{\rm 8, Q_2}=0.794$ and it is based on $\sigma_{8, \Lambda}=0.811$ (as
indicated also in Table~\ref{tab:results}), based on the WMAP7 results~\cite{komatsu08}, but also
consistent with an average over a variety of recent measurements
(see also the corresponding discussion in
 \cite{BPL10}). For the same model but with $\Omega_m=0.3$ we obtain
$\sigma_{\rm 8, Q_2}=0.847$.

\begin{table*}[ht]
\tabcolsep 10pt
\caption[]{Numerical results: The $1^{\rm st}$ column indicates the
  cosmological model. The next three columns list the
corresponding $\Omega_m$, $\sigma_{8}$ and $\delta_{c}$ values, respectively.
The remaining columns present the fractional relative difference of
the abundance of halos between the
Liouville and the $\Lambda$CDM cosmology for two future
cluster surveys discussed in the text and in various redshift intervals.
Error bars are 2$\sigma$ Poisson
uncertainties and are shown only if they are larger than $10^{-2}$.}\label{tab:results}
\vspace{1mm}
\begin{tabular}{|lccc|cc| ccc|} \hline \hline
Model& $\Omega_m$  & $\sigma_{8}$ & $\delta_{c}$ &
\multicolumn{2}{|c}{($\delta{\cal N}/{\cal N}_{\Lambda})_{\rm eROSITA}$} &
\multicolumn{3}{|c|}{$(\delta {\cal N}/{\cal N}_{\Lambda})_{\rm SPT}$} \\
   &  &  &      & $z<0.3$& $0.6\le z <0.9$ & $z<0.3$ & $0.6\le z <0.9$ & $1.2\le z <2$ \\ \hline
$\Lambda$CDM    &0.28  & 0.811  & 1.675 &  \m0.00 & \m0.00           & \m0.00  & \m0.00  & \m0.00 \\
Liouville $Q_2$ &0.28  & 0.794  & 1.676 & $-0.05$ & $-0.27\pm 0.01$&$-0.06$& $-0.15$  & $-0.37$ \\ \hline

$\Lambda$CDM    &0.30  & 0.845  & 1.675 &  \m0.00 & \m0.00           & \m0.00  & \m0.00  & \m0.00 \\
Liouville $Q_2$ &0.30  & 0.847  & 1.674 &  \m0.00 & $-0.16\pm 0.01$ &\m0.00 &$-0.02$  & $-0.26$ \\ \hline \hline 
\end{tabular}
\end{table*}

Given the halo mass function from Eq.~(\ref{MF}) we can now derive an observable
quantity which is the redshift distribution of clusters, ${\cal
N}(z)$, within some determined mass range, say $M_1\le
M/h^{-1}M_{\odot}\le M_2=10^{16}$. This can be estimated by integrating over mass
the expected differential halo mass function, $n(M,z)$:
\be {\cal
N}(z)=\frac{dV}{dz}\;\int_{M_{1}}^{M_{2}} n(M,z)dM,
\ee
where $dV/dz$ is the comoving volume element
\be
\frac{dV}{dz}=\Omega_{s}r^{2}(z)\frac{dr}{dz}
\ee
and $\Omega_{s}$ is the solid angle.
In order to derive observationally relevant
cluster redshift distributions and therefore test the possibility
of discriminating between the Liouville and the $\Lambda$CDM
cosmological models, we will use the expectations of two realistic
future cluster surveys, extensively used in recent literature:
\begin{enumerate}[(a)]
\item the {\tt eROSITA} satellite X-ray survey, with a flux
limit of $f_{\rm lim}=3.3\times 10^{-14}$ ergs s$^{-1}$ cm$^{-2}$,
at the energy band $0.5-5~{\rm keV}$ and covering $\Omega_{s}\sim 20000$ deg$^{2}$ of
the sky;
\item the South Pole Telescope SZ survey, with a limiting
flux density at $\nu_0=150$ GHz of $f_{\nu_0, {\rm lim}}=5$~mJy and
a sky coverage of $\Omega_{s}\sim 4000$~deg$^{2}$.
\end{enumerate}
 
To realize the predictions of the first survey we use the relation
between halo mass and bolometric X-ray luminosity, as a function of
redshift, given in Ref.~\cite{Fedeli}:
\be\label{bolom}
L(M,z)=3.087 \times 10^{44} \left[\frac{M E(z)}{10^{15} h^{-1}
    M_{\odot}} \right]^{1.554} h^{-2} \; {\rm erg s^{-1}} \;.
\ee
The limiting halo mass that can be observed at redshift $z$ is
then found by inserting in the above equation the limiting
luminosity, given by $L=4 \pi d_L^2 f_{\rm lim}${\em c}$_b$, with
$d_L$ the luminosity distance corresponding to the redshift $z$ and
{\em c}$_b$ the band correction, necessary to convert the bolometric
luminosity of Eq.~(\ref{bolom}) to the $0.5-5~{\rm keV}$ band of {\tt
eROSITA}. We estimate this correction by assuming a Raymond-Smith
(1977) plasma model with a metallicity of 0.4$Z_{\odot}$, a typical
cluster temperature of $\sim 4$~keV and a Galactic absorption column
density of $n_{H}=10^{21}$~cm$^{-2}$.

The predictions of the second survey can be realized using again the
relation between limiting flux and halo mass obtained in Ref.~\cite{Fedeli}:
\be\label{sz} f_{\nu_0, {\rm lim}}= \frac{2.592 \times 10^{8} {\rm
mJy}}{d_{A}^{2}(z)} \left(\frac{M}{10^{15} h^{-1}M_{\odot}}\right)^{1.876}
E^{2/3}(z) \; ,\ee 
where $d_A(z) \equiv d_L/(1+z)^2$ is the angular
diameter distance out to redshift $z$.

\begin{figure}[ht] 
\includegraphics[width=0.46\textwidth]{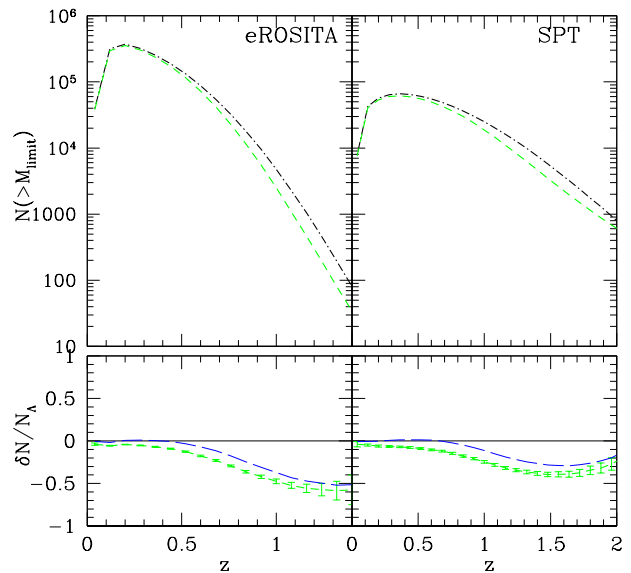}
\caption{The expected
cluster redshift distribution of the Liouville $Q_2$ model
and $\Lambda$CDM (dot-dashed curve) models for $\Omega_m=0.28$
 and for the case of two future cluster surveys (upper panels).
The lower panels show the corresponding fractional difference with respect to the reference
$\Lambda$CDM model (short-dashed curve), while the corresponding curve
for the case of $\Omega_m=0.3$ (see Table~\ref{tab:results}) is shown as the
long-dashed curve. Error bars are 2$\sigma$ Poisson uncertainties.}\label{fig:cluster}
\end{figure}

Below we shall provide predictions for the Liouville cosmology
only for the $Q_{2}$ model (see Table~\ref{tab:chi}). For the other Liouville
models we find that our results remain practically the same (due to the small
differences in the corresponding Hubble flows).
In Fig.~\ref{fig:cluster} (upper panels) we present the expected redshift
distributions above a limiting halo mass, which is $M_1 \equiv M_{\rm
limit}=\max[10^{13.4} h^{-1}M_{\odot}, M_f]$, with $M_f$ corresponding to
the mass related to the flux-limit at the different redshifts,
estimated by solving Eq.~(\ref{bolom}) and Eq.~(\ref{sz}) for $M$. In
the lower panels we present the fractional difference between the
Liouville model and $\Lambda$CDM.
The error-bars shown correspond to 2$\sigma$ Poisson
uncertainties, which however do not include cosmic variance
and possible observational systematic uncertainties, that would
further increase the relevant variance.

The results (see also Table~\ref{tab:results}) indicate that the redshift variation
of the differences between the $Q$ cosmology and $\Lambda$CDM model is only slightly affected by variations in
the value of $\Omega_m$. For
the best-fit case of $\Omega_m=0.28$ we find that significant model differences should be
expected for $z\gtrsim 0.6$, with the
Louville model abundance predictions being always less than those of the
corresponding $\Lambda$CDM model.
Therefore, we have verified that there are observational signatures
that can be used to differentiate the  Liouville model from the
$\Lambda$CDM and possibly from a large class of DE models (see
\cite{BPL10}).
In Table~\ref{tab:results}, one may see a more compact presentation
of our results including the
relative fractional difference between the Liouville model and the
$\Lambda$CDM model, in characteristic redshift bins and for both
future surveys.

\section{Conclusions}\label{sec:conclu}
In this paper, we have studied the
overall dynamics of the Liouville cosmological model
in which the dark energy component is associated with the
dilaton.
In this framework, we first performed a joint
likelihood analysis in order to put tight constraints on the main
cosmological parameters by using the current observational data
[SNIa, BAOs and $H(z)$].
We found that the above models can
accommodate a late-time accelerated expansion.

Secondly, we performed an additional detailed statistical analysis
based on the observed growth rate of clustering and
found that the predicted growth rate of various
Liouville cosmological models
match well the observed growth rate.

The redshift-dependence of the halo abundances, predicted by the
  Liouville model, is significantly different from that of
  the flat $\Lambda$CDM cosmology, at relatively high redshifts
  ($z\gtrsim 0.6$). Future cluster of galaxy surveys can therefore be
  used to discriminate the Liouville model from other contender DE
  models.

\section*{Acknowledgments}

S.B.\ wishes to thank the Dept.\ ECM of the
University of Barcelona for hospitality, and acknowledges financial support from the
Spanish Ministry of Education, within the program of Estancias de
Profesores e Investigadores Extranjeros en Centros Espa\~noles (SAB2010-0118).
The work of N.E.M.\ is supported partly by the London
Centre for Terauniverse Studies (LCTS), using funding from the European
Research Council via the Advanced Investigator Grant 267352.
V.A.M.\ acknowledges support by the Spanish Ministry of Science and Innovation (MICINN) under the project FPA2009-13234-C04-01, by the Spanish Agency of International Cooperation for Development under the PCI project A/030322/10 and by the grant UV-INV-EPDI11-42955 of the University of Valencia.

\appendix
\section{Matter density perturbations in dark energy cosmologies}\label{app:perturb}

As discussed in the text, the Liouville 
cosmology, explored in this paper, is close to GR in the matter dominated era 
(see equation (\ref{friedmann}) and \cite{diamand}). Hence, in this appendix 
we would like to remind the reader, for completeness, 
of some basic elements of 
the equations that govern the evolution of
the mass density contrast, modelled as an ideal fluid 
in the framework of non-interacting 
DE cosmologies that we employ in this work. Also we assume here that 
for small scales (much smaller than the horizon)
the (effective) dark energy component 
is expected to be smooth and thus we 
consider perturbations only on the matter component of the
cosmic fluid. 

In what follows we will adopt a non-relativistic description, based on 
a continuity equation, together with the Euler and
Poisson equations. These equations are:
\be 
\label{cons} 
\left(\frac{\partial \rho}{\partial t}
\right)_{r}+{\bf \nabla}_{r} \cdot (\rho {\bf u})= 0
\ee
\be \label{euler} \left(\frac{\partial {\bf u}}{\partial t}
\right)_{r}+({\bf u} \cdot {\bf \nabla}_{r}) {\bf u}=-{\bf \nabla}
\Phi, 
\ee 
and 
\be \label{poiss} \nabla^{2}_{r}\Phi=  4\pi G \rho
+4\pi G \rho_{DE}(3w_{DE}+1), 
\ee 
where $({\bf r},t)$ are the proper coordinates,
${\bf u}$ is the velocity of a fluid element of volume, $\rho$ is
the mass density, $\rho_{DE}$ is the DE density, $w_{DE}$ is 
the EoS parameter (given in our case by Eq.~(\ref{eos221})) 
and $\Phi$ is the gravitational potential. 
Note that we use here $P_{m}=0$ and $P_{DE}=w_{DE}\rho_{DE}$.

 Upon changing variables from proper $({\bf r},t)$ to comoving $({\bf
x},t)$ ones, the fluid velocity becomes 
\be
\label{vel} {\bf u}=\dot{a} {\bf x}+a {\bf \dot {x}}=\dot{a} {\bf
x}+{\bf v}({\bf x},t), 
\ee 
while the corresponding differential
operators take the following form: \be \label{oper1} {\bf
\nabla}_{x} \equiv {\bf \nabla}=a {\bf \nabla}_{r}, \ee and \be
\label{oper2} \left(\frac{\partial }{\partial t} \right)_{x} \equiv
\frac{\partial }{\partial t}= \left(\frac{\partial }{\partial t}
\right)_{r}+H{\bf x}\cdot {\bf \nabla}, 
\ee 
where ${\bf x}={\bf r}/a$
and ${\bf v}({\bf x},t)$ is the peculiar velocity with respect to
the general expansion. Note that the mass density is written as 
\be
\label{mass1} \rho=\rho_{m}(t)[1+\delta_{m}({\bf x},t)]. 
\ee 
In this context, using the standard evolution equation for the energy density of ordinary dust,  
$$
\dot{\rho}_{m}+3H\rho_{m}=0
$$
as well as Eqs.(\ref{oper1}), (\ref{oper2}) 
and neglecting second-order terms ($\delta_{m} \ll 1$ and $v \ll u$) we 
rewrite Eqs.(\ref{cons}), (\ref{euler}) and (\ref{poiss}) as: 
\be 
\ddot{a} {\bf x}+\frac{\partial {\bf v}}{\partial t}+H
{\bf v}= -\frac{{\bf \nabla}  \Phi}{a}, \ee \be \label{avell} {\bf
{\nabla} \cdot v}=-a \frac{\partial \delta_{m}}{\partial
t}\;, 
\ee 
\be
\frac{1}{a^{2}}\nabla^{2} \Phi=4\pi G
\rho_{m}(1+\delta_{m})+4\pi G(3w_{DE}+1) \rho_{DE}.
\ee 
Following the notation of \cite{Peeb93},
we can write the gravitational potential as follows
\be
\label{phil} 
\Phi=\phi({\bf x},t)+\frac{2}{3}\pi G \rho_{m} a^{2}
x^{2}+\frac{2}{3}\pi G (3w_{DE}+1)\rho_{DE} a^{2}x^{2}. 
\ee 
Thus, utilizing the Friedmann equation and
Eqs.(\ref{phil}), we arrive, after some algebra, at 
\be 
\label{vell1} 
\frac{\partial {\bf v}}{\partial
t}+H {\bf v}= -\frac{{\bf \nabla}  \phi}{a}, 
\ee 
and 
\be
\label{poiss2} 
\nabla^{2} \phi=4\pi G a^{2} \rho_{m} \delta_{m}. 
\ee
Finally, by taking the divergence of Eq. (\ref{vell1}) and using
Eqs. (\ref{avell}) and (\ref{poiss2}), we obtain the time evolution
equation for the matter fluctuation field
\begin{equation}
\label{eq:11} \ddot{\delta}_{m}+2H\dot{\delta}_{m}-
4\pi G \rho_{m} \delta_{m}=0 \;.
\end{equation}
In this context,
$\delta_{m}(a) \propto D(a)$, where $D(a)$ is the linear growing mode
(usually scaled to unity at the present time). 
Note that for those cosmological models which 
adhere to general relativity (GR),
the quantity $G$ reduces to the usual Newton's
gravitational constant $G_{N}$, while in the
case of modified gravity models (see \cite{Lue04,Linder2007,
Tsu08,Gann09}),
we have $G=G_{\rm eff}=Y(a)G_{N}$ (with $Y(a) \ne 1$).

A well known approximate solution to Eq.(\ref{fzz221}) is found
by \cite{Linder2007}, where a growth index $\gamma$ was used
to parameterize the linear growing mode for any type of DE with a time
varying equation of state. Specifically, the approximated 
growth factor was defined through 
\be
D(a)={\rm exp} \left[\int_{1}^{a} 
\frac{\Omega_{m}^{\gamma}(u)}{u} du \right]~,
\label{glind} 
\ee
where now we encapsulate any modification to the $G$ 
in the term
$\Omega_{m}(a)=\frac{8\pi G}{3H^{2}}=\frac{\Omega_{m}a^{-3}}{E^{2}(a)}$. 
This completes our discussion on the growth of perturbations due to ordinary 
matter components of the Liouville cosmology explored in this work.

\vspace{0.5cm}


\begin{thebibliography}{plain}


\bibitem{Teg04} M.~Tegmark \textit{et al.}, Astrophys.\ J.\ \textbf{606}, 702, (2004)

\bibitem{Spergel07} D.~N.~Spergel \textit{et al.}, Astrophys.\ J.\ Suplem.\ \textbf{170},
377, (2007)

\bibitem{essence} T.~M.~Davis \textit{et al.}, Astrophys.\ J.\ \textbf{666},
716, (2007)


\bibitem{Kowal08} M.~Kowalski \textit{et al.}, Astrophys.\ J.\ \textbf{686}, 749, (2008)

\bibitem{Hic09} M.~Hicken \textit{et al.}, Astroplys.\ J.\ \textbf{700}, 1097, (2009)

\bibitem{komatsu08} E.~Komatsu \textit{et al.}, Astrophys.\ J.\ Suplem.\ \textbf{180},
330, (2009); G.~Hinshaw \textit{et al.}, Astrophys.\ J.\ Suplem.\ \textbf{180}, 225,
(2009); E.~Komatsu \textit{et al.}, Astrophys.\ J.\ Suplem.\ \textbf{192}, 18, (2011)

\bibitem{LJC09} J.~A.~S.~Lima and  J.~S.~Alcaniz, Mon.\ Not.\ Roy.\
Astron.\ Soc.\ {\bf 317}, 893 (2000) [astro-ph/0005441];
J.~F.~Jesus and J.~V.~Cunha, Astrophys.\ J.\ Lett.\ {\bf 690}, L85
(2009) [arXiv:0709.2195]

\bibitem{BasPli10}
S.~Basilakos and M.~Plionis, Astrophys.\ J.\ Lett.\ {\bf 714}, 185 (2010)

\bibitem{Weinberg89} S.~Weinberg, Rev.\ Mod.\ Phys.\ \textbf{61}, 1, (1989)

\bibitem{Peebles03} P.~J.~Peebles and B.~Ratra, Rev.\ Mod.\ Phys.\ \textbf{75},
559, (2003)

\bibitem{Pad03} T.~Padmanabhan, Phys.\ Rept.\ \textbf{380}, 235, (2003)

\bibitem{Zlatev99} I.~Zlatev, L.~Wang, and P.~J.~Steinhardt, Phys.\ Rev.\ Lett.\ {\bf 82},
896 (1999); L.~P.~Chimento, A.~S.~Jakubi, D.~Pavon, and W.~Zimdahl,
Phys.\ Rev.\ D {\bf 67}, 083513 (2003); S.~Nojiri and S.~D.~Odintsov,
Phys.\ Lett.\ B {\bf 637}, 139 (2006); M.~Quartin, M.~O.~Calv\~ao, S.~E.~Joras, R.~R.~R.~Reis, and I.~Waga, JCAP {\bf 05}, 007 (2008);
P.~J.~Steinhardt, in: \textit{Critical Problems in
Physics}, edited by V.~L.~Fitch, D.~R.~Marlow and M.~A.~E.~Dementi (Princeton
Univ.\ Pr., Princeton, 1997); P.~J.~Steinhardt, Phil.\ Trans.\ Roy.\ Soc.\ Lond.\
\textbf{A361}, 2497, (2003)

\bibitem{AASW09} S.~del~Campo, R.~Herrera, and D.~Pavon, Phys.\ Rev.\ D{\bf 78},
021302 (2008); E.~Abdalla, L.~R.~Abramo, and J.~C.~C.~de~Souza,
Phys.\ Rev.\ D {\bf 82}, 023508 (2010); S.~Z.~W.~Lip, Phys.\ Rev.\ D
{\bf 83}, 023528 (2011) 

\bibitem {Egan08} C.~A.~Egan and C.~H.~Lineweaver, Phys.\ Rev.\ D  \textbf{78},
3528, (2008)

\bibitem {Dolgov82} A.~D.~Dolgov, in: \textit{The very Early Universe}, Ed.\
G.~Gibbons, S.~W.~Hawking, S.~ T.~Tiklos (Cambridge U., 1982)

\bibitem {Caldwell98} R.~R.~Caldwell, R.~Dave and P.~J.~Steinhardt, Phys.\ Rev.\
Lett.\ \textbf{80}, 1582 (1998)

\bibitem {Jassal} H.~K.~Jassal, J.~S.~Bagla, T.~Padmanabhan, Phys.\ Rev.\ D
\textbf{72} 103503 (2005); H.~K.~Jassal, J.~S.~Bagla, T.~Padmanabhan, Mon.\
Not.\ Roy.\ Astron.\ Soc.\ Letters  \textbf{356}, L11-L16, (2005)

\bibitem {SR} L.~Samushia, B.~Ratra, Astrophys.\ J.\ \textbf{650}, L5 (2006);
Astrophys.\ J.\ \textbf{680}, L1 (2008)

\bibitem {Xin} J.~Q.~Xia, H.~Li, G.~B.~Zhao and X.~Zhang, Phys.\ Rev.\ D
\textbf{78}, {083524} {(2008)}; G.B. Zhao, J.Q. Xia, B. Feng, X. Zhang, Int.
J. Mod.Phys. \textbf{D16} 1229, (2007); J.~Q.~Xia, G.~B.~Zhao, B.~Feng, H.~Li,
X.~Zhang, Phys.\ Rev.\ D \textbf{73}, 063521 (2006)

\bibitem {SVJ} J.~Simon, L.~Verde, R.~Jim\'enez, Phys.\ Rev.\ D \textbf{71},
123001 (2005)

\bibitem{phantom} V.~Faraoni, Int.\ J.\ Mod.\ Phys.\ D {\bf 11}, 471 (2002); 
R.~R.~Caldwell, M.~Kamionkowski, and N.~N.~Weinberg, Phys.\ Rev.\ Lett.\ {\bf 91}, 071301
(2003); J.~A.~S.~Lima and J.~S.~Alcaniz, Phys.\ Lett.\ B \textbf{600}, 191
(2004) [astro-ph/0402265]; G.~Izquierdo and D.~Pavon, Phys.\ Lett.\ B
\textbf{639}, 1 (2006) [arXiv:gr-qc/0606014]; S.~H.~Pereira and J.~A.~S.~Lima, Phys.\ Lett.\ B \textbf{669}, 266 (2008) [arXiv:0806.0682 [astro-ph]]


\bibitem {Mota04} D.~F.~Mota and C.~van~de~Bruck, Astronomy \& Astrophysics,
\textbf{421}, 71 (2004)

\bibitem {Nunes06} N.~J.~Nunes and D.~F.~Mota, Mon.\ Not.\ Roy.\ Astron.\ Soc.\
\textbf{368}, 751 (2006)

\bibitem {Basi09} S.~Basilakos, J.~C.~Sanchez, L.~Perivolaropoulos, Phys.\ Rev.\
D \textbf{80}, 3530 (2009)


\bibitem {Ame10} E.~J.~Copeland, M.~Sami and S.~Tsujikawa, Intern.\
Journal of Modern Physics D \textbf{15}, 1753 (2006); L.~Amendola and S.~Tsujikawa, ``Dark Energy Theory and Observations,'' Cambridge University Press,
Cambridge UK (2010); R.~R.~Caldwell and M.~Kamionkowski,
Ann.\ Rev.\ Nucl.\ Part.\ Sci.\ \textbf{59}, 397 (2009) [arXiv:0903.0866]

\bibitem{Gas}
M.~Gasperini and G.~Veneziano, Phys.\ Rept.\ {\bf 373}, 1 (2003);
J.~E.~Lidsey, D.~Wands and E.~J.~Copeland,  Phys.\ Rept.\ {\bf 337}, 343 (2000)


\bibitem{emnw} J.~R.~Ellis, N.~E.~Mavromatos and D.~V.~Nanopoulos,
  Phys.\ Lett.\ B {\bf 619} (2005) 17 [arXiv:hep-th/0412240];
J.~R.~Ellis, N.~E.~Mavromatos, D.~V.~Nanopoulos and M.~Westmuckett,
  Int.\ J.\ Mod.\ Phys.\  A {\bf 21} (2006) 1379 [arXiv:gr-qc/0508105].

\bibitem{diamandis}
  G.~A.~Diamandis, B.~C.~Georgalas, N.~E.~Mavromatos and E.~Papantonopoulos,
  Int.\ J.\ Mod.\ Phys.\ A {\bf 17} (2002) 4567 [arXiv:hep-th/0203241];
G.~A.~Diamandis, B.~C.~Georgalas, N.~E.~Mavromatos, E.~Papantonopoulos and
I.~Pappa,
  Int.\ J.\ Mod.\ Phys.\ A {\bf 17} (2002) 2241 [arXiv:hep-th/0107124].

\bibitem{diamandis2} G.~A.~Diamandis, B.~C.~Georgalas, A.~B.~Lahanas, N.~E.~Mavromatos and D.~V.~Nanopoulos,
  Phys.\ Lett.\  B {\bf 642} (2006) 179 [arXiv:hep-th/0605181].

\bibitem{lmn} A.~B.~Lahanas, N.~E.~Mavromatos and D.~V.~Nanopoulos,
  Phys.\ Lett.\  B {\bf 649} (2007) 83 [arXiv:hep-ph/0612152];
  B.~Dutta, A.~Gurrola, T.~Kamon, A.~Krislock, A.~B.~Lahanas, N.~E.~Mavromatos, D.~V.~Nanopoulos,
  Phys.\ Rev.\  {\bf D79}, 055002 (2009).
  [arXiv:0808.1372 [hep-ph]].

\bibitem{aben} I.~Antoniadis, C.~Bachas, J.~R.~Ellis and D.~V.~Nanopoulos,
  Nucl.\ Phys.\ B {\bf 328} (1989) 117;
  Phys.\ Lett.\ B {\bf 257} (1991) 278.

\bibitem{ddk} F.~David,
  Mod.\ Phys.\ Lett.\ A {\bf 3} (1988) 1651;
J.~Distler and H.~Kawai,
  Nucl.\ Phys.\ B {\bf 321} (1989) 509;
J.~Distler, Z.~Hlousek and H.~Kawai,
  Int.\ J.\ Mod.\ Phys.\ A {\bf 5} (1990) 391;
see also: N.~E.~Mavromatos and J.~L.~Miramontes,
  Mod.\ Phys.\ Lett.\ A {\bf 4} (1989) 1847;
E.~D'Hoker and P.~S.~Kurzepa,
  Mod.\ Phys.\ Lett.\ A {\bf 5} (1990) 1411.

\bibitem{emn} J.~R.~Ellis, N.~E.~Mavromatos and D.~V.~Nanopoulos,
  Phys.\ Lett.\ B {\bf 293} (1992) 37 [arXiv:hep-th/9207103];
  Mod.\ Phys.\ Lett.\ A {\bf 10} (1995) 1685 [arXiv:hep-th/9503162];
  Chaos Solitons Fractals {\bf 10} (1999) 345 [arXiv:hep-th/9805120].
  
\bibitem{lahanas2010} A.~B.~Lahanas,
  Phys.\ Rev.\  {\bf D83}, 103523 (2011)
  [arXiv:1102.4277 [hep-ph]].

\bibitem{mitsou} J.~R.~Ellis, N.~E.~Mavromatos, V.~A.~Mitsou and D.~V.~Nanopoulos,
  Astropart.\ Phys.\  {\bf 27}, 185 (2007)
  [arXiv:astro-ph/0604272].

\bibitem{mitsou2} N.~E.~Mavromatos and V.~A.~Mitsou,
  Astropart.\ Phys.\  {\bf 29}, 442 (2008)
  [arXiv:0707.4671 [astro-ph]].


\bibitem{Stern10}
D.~Stern, R.~Jimenez, L.~Verde, M.~Kamionkowski, S.~A.~Stanford,
J.\ Cosmol.\ Astropart.\ Phys.\ {\bf 02}, 008 (2010)

\bibitem{Eis05}
D.~J.~Eisenstein \textit{et al.}, Astrophys.\ J.\ {\bf 633}, 560 (2005); 
N.~Padmanabhan \textit{et al.}, Mon.\ Not.\ Roy.\ Astron.\ Soc.\ {\bf 378}, 852
(2007).

\bibitem{Perc10} W.~Percival \textit{et al.},
Mon.\ Not.\ Roy.\ Astron.\ Soc.\ {\bf 401}, 2148 (2010).

\bibitem{Kazin10}
E.~A.~Kazin, Astrophys.\ J.\ {\bf 710}, 1444 (2010).

\bibitem{press}  W.~H.~Press and P.~Schechter,
  Astrophys.\ J.\  {\bf 187}, 425 (1974)

\bibitem{Reed} D.~Reed, R.~Bower, C.~Frenk, A.~Jenkins, and T.~Theuns, Mon.\ Not.\ Roy.\ Astron.\ Soc.\
{\bf 374}, 2 (2007).

\bibitem{Evra}
A.~E.~Evrard \textit{et al.}, Astrophys.\ J.\ {\bf 573}, 7 (2002).

\bibitem{Borg01}
S.~Borgani \textit{et al.}, Astrophys.\ J.\ {\bf 561}, 13 (2001).

\bibitem{Reip02} T.~H.~Reiprich, H.~B\"ohringer, Astrophys.\ J.\ {\bf 567}, 716 (2002).

\bibitem{Vik09}
A.~Vikhlinin \textit{et al.}, Astrophys.\ J.\ {\bf 692}, 1060 (2009).

\bibitem{Bat98}
M.~Bartelmann, A.~Huss, J.~M.~Colberg, A.~Jenkins, and F.~R.~Pearce,
Astron.\ Astrophys.\ {\bf 330}, 1 (1998).

\bibitem{Dahle06} H.~Dahle, Astrophys.\ J.\ {\bf 653}, 954 (2006).

\bibitem{Corl09} V.~L.~Corless, L.~J.~King, Mon.\ Not.\ Roy.\ Astron.\ Soc.\ {\bf 396}, 315 (2009).

\bibitem{Bah03} N.~A.~Bahcall \textit{et al.}, Astrophys.\ J.\ {\bf 585}, 182 (2003).

\bibitem{Wen10} Z.~L.~Wen, J.~L.~Han, F.~S.~Liu, Mon.\ Not.\ Roy.\ Astron.\ Soc.\ {\bf 407}, 553 (2010)

\bibitem{Taub05}
J.~A.~Tauber, New Cosmological Data and the values of the
Fundamental Parameters, {\bf 201}, 86 (2005).

\bibitem{Wein03}
N.~N.~Weinberg  and  M.~Kamionkowski, Mon.\ Not.\ Roy.\ Astron.\ Soc.\ {\bf 341}, 251
(2003).

\bibitem{Liberato}
L.~Liberato and R.~Rosenfeld, JCAP {\bf 0607}, 009 (2006).

\bibitem{manera}
M.~Manera and D.~F.~Mota, Mon.\ Not.\ Roy.\ Astron.\ Soc.\  {\bf 371},
1373 (2006).

\bibitem{Abramo07}
L.~R.~Abramo, R.~C.~Batista, L.~Liberato, and R.~Rosenfeld, JCAP
{\bf 11}, 012 (2007).


\bibitem{Fran08}
M.~J.~Francis, G.~F.~Lewis, and E.~V.~Linder, Mon.\ Not.\ Roy.\ Astron.\
Soc.\  {\bf 393}, L31 (2009); M.~J.~Francis, G.~F.~Lewis and E.~V.~Linder, Mon.\ Not.\ Roy.\ Astron.\ Soc.\ (2009).

\bibitem{Sch09}
F.~Schmidt, A.~Vikhlinin and Wayne~Hu, Phys.\ Rev.\ D {\bf 80}, 083505 (2009)

\bibitem{Mort09}
M.~J.~Mortonson, Phys.\ Rev.\ D {\bf 80}, 123504 (2009).

\bibitem{Rap10}	
D.~Rapetti, S.~W.~Allen, A.~Mantz and H.~Ebeling, Mon.\ Not.\ Roy.\ Soc.\ {\bf 406}, 1796 (2010)

\bibitem{Pace10}
F.~Pace, J.-C.~Waizmannm and M.~Bartelman, Mon.\ Not.\ Roy.\ Astron.\ Soc.\
{\bf 406}, 1865 (2010).

\bibitem{Alam10}
U.~Alam, Z.~Lukic\' and S.~Bhattacharya,  Astrophys.\ J.\ {\bf 727}, 87 (2011)


\bibitem{Khed10}
S.~Khedekar, and S.~Majumdar, Phys.\ Rev.\ D {\bf 82}, 081301 (2010);
S.~Khedekar, S.~Majumdar and S.~Das, Phys.\ Rev.\ D {\bf 82}, 041301 (2010).

\bibitem{BPL10}
S.~Basilakos, M.~Plionis, A.~Lima, Phys.\ Rev.\ D {\bf 82}, 083517  (2010)

\bibitem{Lomb10}
L.~Lombriser, A.~Slosar, U.~Seljak and Wayne~Hu, arXiv:1003.3009 (2010)


\bibitem{diamand}
  G.~A.~Diamandis, B.~C.~Georgalas, A.~B.~Lahanas, N.~E.~Mavromatos and D.~V.~Nanopoulos,
  Phys.\ Lett.\  B {\bf 642}, 179 (2006)
  [arXiv:hep-th/0605181].

\bibitem{Sah02}
V. Sahni, T. D. Saini, A. A. Starobinsky and U. Alam, JETP Lett., {\bf 77}, 
201 (2003)

\bibitem{Riess10}
A. G. Riess, et al., Astrophys.\ J.\ \textbf{730}, 119, (2011); 
Astrophys.\ J.\ \textbf{732}, 129, (2011) 

\bibitem{kaluza}
M.~Minamitsuji, M.~Sasaki and D.~Langlois,
    Phys.\ Rev.\ D {\bf 71}, 084019 (2005)
    [arXiv:gr-qc/0501086].

\bibitem{Saini00}
T.~D.~Saini, S.~Raychaudhury, V.~Sahni, and A.~A.~Starobinsky,
Phys.\ Rev.\ Lett.\ {\bf 85}, 1162 (2000);
D.~Huterer and M.~S.~Turner, Phys.\ Rev.\ D {\bf 64}, 123527 (2001).

\bibitem{Linjen03}
E.~V.~Linder and A.~Jenkins, Mon.\ Not.\ Roy.\ Astron.\ Soc.\ {\bf 346},
573 (2003).

\bibitem{Linder05}
E.~V.~Linder, Phys.\ Rev.\ D {\bf 72}, 043529 (2005)

\bibitem{PLC}
M.~Plionis, P.~Coles and P.~Catelan,
Mon.\ Not.\ Roy.\ Astron.\ Soc.\ {\bf 262}, 465 (1993)

\bibitem{San06}
A.~G.~Sanchez, C.~M.~Baugh, W.~J.~Percival, J.~A.~Peacock,
N.~D.~Padilla, S.~Cole, C.~S.~Frenk and P.~Norberg,
Mon.\ Not.\ Roy.\ Astron.\ Soc.\ {\bf 366}, 189 (2006)

\bibitem{Fel03}
H.~Feldman \textit{et al.}, Astrophys.\ J.\ Lett.\ {\bf 596}, L131 (2003)

\bibitem{MoTu05}
R.~Mohayaee and B., Astron.\ J.\ {\bf 130}, 1502 (2005)

\bibitem{And05}
H.~Andernach, M.~Plionis, O.~Lopez-Cruz and E.~Tago,
S.~Basilakos, Astronomical Society of the Pacific
Conference Series {\bf 329}, 289 (2005)

\bibitem{Sch02}
S.~Schindler, Space Science Reviews {\bf 100}, 299 (2002)


\bibitem{Lue04}
A.~Lue, R.~Scossimarro, and G.~D.~Starkman, Phys.\ Rev.\ D {\bf 69}, 124015 (2004)

\bibitem {Stab06}
F.~H.~Stabenau and B.~Jain, Phys.\ Rev.\ D {\bf 74}, 084007 (2006)

\bibitem {Uzan07}
P.~J.~Uzan, Gen.\ Rel.\ Grav.\ {\bf 39}, 307 (2007)

\bibitem{Linder2007}
E.~V.~Linder, Phys.\ Rev.\ Lett.\ {\bf 70}, 023511 (2004);
E.~V.~Linder and R.~N.~Cahn, Astrop.\ Phys.\ {\bf 28}, 481 (2007);
D. Huterer and E.~V.~Linder, Phys.\ Rev.\ D.\ {\bf 75}, 023519 (2007)
 
\bibitem {Tsu08}
S.~Tsujikawa, K.~Uddin and R.~Tavakol, Phys.\ Rev.\ D {\bf 77}, 043007 (2008)


\bibitem {Gann09}
R.~Gannouji, B.~Moraes and D.~Polarski, 
J.\ Cosmol.\ Astropart.\ Phys.\ {\bf 02}, 034 (2009)


\bibitem{Dave02}
R. Dave, R. R. Caldwell and P. J. Steinhardt, Phys. Rev. D., {\bf 66}, 023516 (2002)

\bibitem{Boi00}
B. Boisseau, G. Esposito Farese, D. Polarski and A.A. Starobinski,
Phys. Rev. Lett., {\bf 85} 2236, (2000) 


\bibitem{Wang98}
L.~Wang and J.~P.~Steinhardt, Astrophys.\ J.\ {\bf 508}, 483 (1998).

\bibitem{Peeb93}
P.~J.~E.~Peebles, ``Principles of Physical Cosmology'', Princeton
University Press, Princeton New Jersey (1993).


\bibitem{Nes08}
  S.~Nesseris and  L.~Perivolaropoulos,
  Phys.\ Rev.\ D {\bf 77}, 023504 (2008)

\bibitem{Blake} 
C.~Blake et al., Mon. Not. Roy. Astron. Soc., 
{\bf 415}, 2876 (2011)

\bibitem{Sam11}
L. Samushia, W. J. Percival and A. Raccanelli, [arXiv:1102.1014], (2012) 

\bibitem{Simp10}
F., Simpson and J. A., Peacock, Phys. Rev. D., {\bf 81}, 043512 (2010)

\bibitem{Wei}
H. Wei, Phys. Lett. B., {\bf 664}, 1 (2008); 
Y. Gong, Phys. Rev. D., {\bf 78}, 123010 (2008); 
J. Dosset, et al., JCAP, {\bf 1004}, 022 (2010)

\bibitem{Zhang}
Wen-Shuai Zhang, et al., arXiv:1202.0892 (2012) 

\bibitem{DGP}
G. Dvali, G. Gabadadze and M. Porrati, Phys. Lett. B., {\bf 485}, 208, (2000)


\bibitem{Pol}
D.~Polarski and R.~Gannouji, Phys.\ Lett.\ B {\bf 660}, 439 (2008)

\bibitem{Bas11}
J.~F.~Jesus, F.~A.~Oliveira, S.~Basilakos and J.~A.~S.~Lima, 
arXiv:1105.1027 (2011)

\bibitem{McDon05}
P.~McDonald \textit{et al.}, Astrophys.\ J.\ {\bf 635}, 761 (2005).


\bibitem{Guzzo08}
L.~Guzzo {\it et al.}, Nature {\bf 451}, 541 (2008)

\bibitem{Verde02}
L. Verde, et al., Mon. Not. Roy. Astron. Soc., {\bf 335}, 432, (2002)

\bibitem{Hawk03}
E. Hawkins, et al., Mon. Not. Roy. Astron. Soc., {\bf 346}, 78, (2003)


\bibitem{Teg06}
M.~Tegmark {\it et al.}, Phys.\ Rev.\ D {\bf 74}, 123507 (2006).

\bibitem{Ross07}
N.~P.~Ross {\it et al.}, Mon.\ Not.\ Roy.\ Astron.\ Soc.\ {\bf 381}, 573
(2007).

\bibitem{daAng08}
J.~da~Angela {\it et al.}, Mon.\ Not.\ Roy.\ Astron.\ Soc.\ {\bf 383}, 565
(2008)

\bibitem{Viel04}
M.~de~Viel, M.~G.~Haehnelt and V.~Springel, Mon.\ Not.\ Roy.\ Astron.\ Soc.\
{\bf 354}, 684 (2004); Y. Gong, Phys.\ Rev.\ D {\bf 78}, 123010 (2008)


\bibitem{peacock} See, for instance: J.~A.~Peacock, \emph{Cosmological Physics} (Cambridge Univ.\ Press, Cambridge (UK),  2003).

\bibitem{eke}
V.~Eke, S.~Cole, C.~S.~Frenk, Mon.\ Not.\ Roy.\ Astron.\ Soc.\ {\bf 282}, 263 (1996)

\bibitem{Bard86} J.~M.~Bardeen, J.~R.~Bond, N.~Kaiser, and A.~S.~Szalay,
Astrophys.\ J.\ {\bf 304}, 15 (1986);
N.~Sugiyama, Astrophys.\ J.\ Suplem.\ {\bf 100}, 281 (1995) 

\bibitem{Jenk01} A. Jenkins, \textit{et al.}, Mon. Not. Roy. Astron. Soc., {\bf 321}, 372 (2001).

\bibitem{LM07} L.~Marassi and J.~A.~S.~Lima, Int.\ J.\ Mod.\ Phys.\ D {\bf 13}, 1345 (2004); {\it ibid}.\ {\bf 16}, 445 (2007).

\bibitem{Fedeli} C.~Fedeli, L.~Moscardini and S.~Matarrese,
Mon.\ Not.\ Roy.\ Astron.\ Soc.\
{\bf 397}, 1125 (2009)



\end{thebibliography}
\end{document}